\documentclass[11pt]{aastex}
\usepackage{graphicx,emulateapj5,apjfonts}
\usepackage{onecolfloat}
\usepackage{epsf}
\usepackage{natbib}
\usepackage{graphicx}
\usepackage{xspace}
\usepackage{longtable,lscape}
\bibpunct {(} {)} {;} {a} {} {,}

\shortauthors{Brasseur et al.}
\shorttitle{Fiducial Stellar Population Sequences for the $VJK_S$ Photometric System}

\newcommand\Teff{{T_{\rm eff}}}
\accepted{}

\slugcomment{{\sc The Astronomical Journal, accepted }  }

\begin{document}


\def\head{

\title{Fiducial Stellar Population Sequences for the $VJK_S$ Photometric System}

\author{
Crystal M.~Brasseur\altaffilmark{1}, 
Peter B.~Stetson\altaffilmark{2},
Don A.~VandenBerg\altaffilmark{1}, 
Luca Casagrande\altaffilmark{3}, 
Giuseppe Bono\altaffilmark{4}, 
Massimo Dall'Ora \altaffilmark{5}}
 \begin{center}
{\scriptsize

$^{1}${Department of Physics \& Astronomy, University of Victoria, P.O.~Box 3055, Victoria, B.C., V8W~3P6, Canada}

$^{2}${Herzberg Institute of Astrophysics, National Research Council Canada, 5071 W.~Saanich Rd., Victoria, B.C., V9E~2E7, Canada}

$^{3}${Max-Planck-Institut f\"ur Astrophysik, Postfach 1317, 85748 Garching, Germany}

$^{4}${Universita' di Roma Tor Vergata, Via della Ricerca Scientifica 1, 00133 Rome, Italy}

$^{5}${INAF-Osservatorio Astronomico di Capodimonte, Via Moiariello 16, 80131 Napoli, Italy}

}
 \end{center}

\email{brasseur@mpia.de}

\begin{abstract}

We have obtained broad-band near-infrared photometry for seven
Galactic star clusters (M$\,$92, M$\,$15, M$\,$13, M$\,$5,
NGC$\,$1851, M$\,$71 and NGC$\,$6791) using the WIRCam wide-field
imager on the Canada-France-Hawaii Telescope, supplemented by images
of NGC$\,$1851 taken with HAWK-I on the VLT.  In addition, 2MASS
observations of the [Fe/H] $\approx 0.0$ open cluster M$\,$67 were
added to the cluster database.  From the resultant $(V-J)$-$V$
and $(V-K_S)$-$V$ colour-magnitude diagrams (CMDs), fiducial sequences
spanning the range in metallicity, {$-2.4 \lesssim \textnormal{[Fe/H]}
\lesssim +0.3$}, have been defined which extend (for most clusters)
from the tip of the red-giant branch (RGB) to $\sim 2.5$ magnitudes below
the main-sequence turnoff.  These fiducials provide a valuable set of
empirical isochrones for the interpretation of stellar population data
in the 2MASS system.  We also compare our newly derived CMDs to Victoria
isochrones that have been transformed to the observed plane using recent
empirical and theoretical colour-${T_{\rm eff}}$ relations.  The models
are able to reproduce the entire CMDs of clusters more metal rich than
[Fe/H] $\approx -1.4$ quite well, on the assumption of the same
reddenings and distance moduli that yield good fits of the same
isochrones to Johnson-Cousins $BV(RI)_C$ photometry.  However, the
predicted giant branches become systematically redder than the observed
RGBs as the cluster metallicity decreases.  Possible explanations for these discrepancies are discussed. 
\end{abstract}

}

\twocolumn[\head]

\section{Introduction}
\label{sec:intro}

The Two Micron All Sky Survey (2MASS) has uniformly scanned the entire sky
in three near-infrared bands, $JHK_S$.  Cataloging more than 300 million
point sources, 2MASS observations are not only advancing our understanding
of stars, but they also provide the photometric standards used to calibrate
all other $JHK_S$ photometry.  In recent years, a new generation of
$JHK_S$ equipped observing facilities have come online, driven by the benefits
of observing at these longer wavelengths.  In particular, reduced dust
attenuation allows the near-infrared to be a better probe of stellar
populations in dust-obscured and heavily reddened galaxies.  Furthermore,
the integrated luminosities of intermediate-age and old stellar populations,
which are predominantly due to their asymptotic-giant-branch (AGB) and
red-giant-branch (RGB) stars, are brightest at these longer wavelengths.

As the infrared (IR) region of the electromagnetic spectrum receives
growing attention in modern astrophysics, it becomes desirable to have
deep near-IR photometry for Galactic star clusters.  Currently, the
available near-IR fiducials of these systems remain restricted in
both metallicity and luminosity, with the observations rarely extending
to fainter magnitudes than the base of the RGB
(e.g., \citealp{F83,V07,F00}).  Star clusters are the ideal stellar
populations to observe because, notwithstanding a growing number of exceptions {(e.g.,
$\omega$ Cen)}, their stars are or can 
be considered homogeneous in both age and initial chemical composition to a
rather good approximation, at least for a number of purposes.  Therefore, observations of these
systems provide us with exceedingly valuable stellar population templates
across a broad range of stellar parameter space, as well as the data which
are needed to test and refine the predicted colours from model atmospheres
and the temperatures given by stellar evolutionary models.

Stellar population studies rely on the accuracy of both colour--$\Teff$
relations and the stellar $\Teff$ scale in order to derive the ages,
metallicities, and star formation histories of stellar systems based on
isochrone fits to observed colour-magnitude diagrams (CMDs).  One way
to test these relations is to obtain photometry in many bandpasses and then
to determine the level of consistency across all possible CMDs (and
colour-colour diagrams) that can be generated.  While considerable work
has been carried out to test and improve the colour transformations for
the $BV(RI)_C$ and Str{\"o}mgren filter systems (e.g., \citealt{VC03}; \citealt{Clem2004}), 
very little has been done to date on the
colour-$\Teff$ relations for the near-IR (two of the few works to date are \citet{Pins2004} and citet{GS2003}).

In this paper, new CMDs for open
and globular star clusters, along with published data for field subdwarfs,
are compared with isochrones in order to assess the reliability of recent
$(V-J)$--$\Teff$ and $(V-K_S)$--$\Teff$ relations.  Additionally, and
perhaps most importantly, we can for the first time evaluate the 
consistency of the isochrone fits across optical and infrared colours.
The companion work to this paper, VandenBerg et al. (AJ, submitted);
hereafter known as VCS10), focusses on the colour--$\Teff$ relations
for the Johnson-Cousins $BV(RI)_C$ photometric system.  In both
papers, two recently developed sets of colour transformations are tested.
The first of these has been derived  (see VCS10) from the latest MARCS model
atmospheres (\citealt{Gust2008}), while the second is the set of empirical
relations developed by \citet{Casa2010} who used the Infrared Flux
Method (IRFM) to produce colour--$\Teff$ relations for dwarf and subgiant
stars spanning a large range in metallicity.

\vspace{1cm}

Addressing the need for precise fiducial sequences in the
near-infrared, we have obtained observations of the seven Galactic star
clusters, M$\,$92, M$\,$15, M$\,$13, M$\,$5, NGC$\,$1851, M$\,$71 and
NGC$\,$6791.  In \textsection \ref{sec:observations}, we describe our
observing programs and present salient details concerning the reduction 
of the data --- including, in particular, the calibration of our
photometry to the standard 2MASS system.  In  \textsection
\ref{sec:Fiducials}, the CMDs and cluster fiducials are presented 
which provide template stellar population sequences for the range in
metallicity, $-2.4\lesssim$ [Fe/H] $\lesssim +0.3$.  An examination of how well isochrones that employ
the synthetic MARCS, and the empirical \citet{Casa2010}, colour--$\Teff$
relations are able to reproduce the observed $VJK_S$ photometry of
the local subdwarfs and of our target clusters is presented in
\textsection \ref{sec:ctrelations}.  Finally, a short summary of our
results, as well as a brief discussion of the usefulness of the
derived fiducial sequences for stellar populations research, is
given in \textsection \ref{sec:summary}.

\begin{figure}[ht]
\includegraphics[bb= 0 140 446 700,  width=7cm]{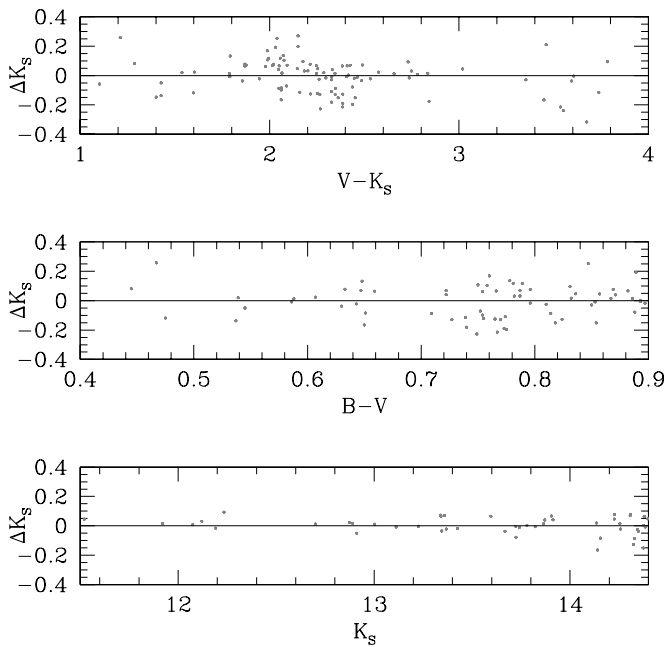}
\caption{$K_S$-band photometric differences for stars observed in both 2MASS and our VLT fields for NGC$\,$1851.}
\label{photometry}
\end{figure}

\section{Observations}
\label{sec:observations}

\begin{table*}
 \hspace{0.033\textwidth}
 \centering
\begin{tabular}{l c c c c c} \hline \hline
Cluster & Dates of Observations & Exposure time in J &Exposure time in K &  Weather Conditions  \\
  \hline
M$\,$15 & Aug 10-18, 2008 &  45s x 40, 5s x 7 & 20s x 97  & photometric   \\
M$\,$92 & Aug 10-18, 2008 &  25s x 5, 45s x 14 , 5s x 21 & 20s x 127, 5s x 19 & photometric   \\
M$\,$13& Aug 11-12, 2008 &  45s x 25 & 20s x 153  &photometric  \\
M$\,$5& June 11, 2009 &  20s x 56  &...&  photometric  \\
NGC 1851 &  Jan 30, 2008  &...& 10s x 139&  photometric\\
M$\,$71 & Aug 10-18, 2008  &  5s x 22 & 5s x 45&  photometric   \\
NGC$\,$6791& Aug 10-18, 2008 &  5s x 21 &... &  photometric   \\

\hline
 \end{tabular}
\caption[Description of observations for each target. ]{Description of observations for each target.}
\label{observing}
\end{table*}

In the 2008B semester, we received WIRCam time on the CFHT to observe five
Galactic star clusters (NGC$\,$6791, M$\,$13, M$\,$15, M$\,$92 and
M$\,$71) in $J$ and $K_S$.  In the following semester (2009A), we obtained
observations of M$\,$5, but only in the $J$ band due to poor weather
conditions. In addition, $K_S$-band images of NGC$\,$1851 that had been
previously collected by two of us (G.B.~and M.D.) using the VLT HAWK-I detector
were included in the dataset.

All of our CFHT observations aimed to reach a signal-to-noise ratio
of 25 at $\sim 2.5$ magnitudes below the main sequence turn-off (MSTO)
of each cluster.  To accomplish this, a series of long exposures, together
with several short-exposure images (which were critical for the calibration
of our fields to 2MASS photometry), were taken in the $J$ and $K_S$ filters
for each cluster on our target list.  By taking a series of exposures,
each star was detected multiple times, thereby helping to improve the
precision of the final photometry. We direct the reader to Table 1 which lists
the image properties of each cluster. Unlike optical photometry, where one would separately observe standard star fields, the 2MASS All-Sky Point Source Catalog contains enough stars in each of our clusters which can be used as standards. For example, M$\,$13 has 3221 stars in the 2MASS All-Sky Point Source Catalog within 30 arcmin of its center. Of these, we selected stars with the lowest claimed photometric errors (typically errors $<$ 0.02) for the calibration of our frames.

In addition to the on-target images, it was necessary to obtain a large
number of off-target images of the sky because both the spatial and
temporal variations of atmospheric emission in the IR are quite
considerable ($\sim 10$\% over a timespan of as little as 10 minutes).
Indeed, particular care was taken in determining how to construct
and subtract a sky image from a target image, since the quality of
a processed image is dominated by precisely how this subtraction is
carried out.  In order to minimize the effects of the variable sky on 
our photometry, we decided to image an equal number of sky frames 
off-target as science frames on-target.  We also chose to employ a large
dither pattern so as to remove bad pixels when median combining the frames.
Once observed, images were pre-processed by CFHT staff using the
WIRCam pipeline.  This included flat fielding, bias and dark
subtraction, as well as sky subtraction.

Upon receiving the
pre-processed images, instrumental magnitudes for all stars were
obtained using the point spread function (PSF) modeling and fitting
techniques in the DAOPHOT/ ALLSTAR/ ALLFRAME packages
(\citealt{Stetson1987}, \citealt{StetsonHarris1988}).  In essence,
these programs work by detecting stars on a specific image, building a
model PSF from a few isolated, bright stars and then subtracting this
PSF from all stars detected.  (For a more detailed description of how
how these programs work, see \citealt{Stetson1987}.)  The standard star
calibration directly to 2MASS stars in the same frame, was done by
solving for the zero point, and colour terms which accounts for the difference in
the central wavelengths of the CFHT and 2MASS $J$, and $K$ filters.

\begin{figure}
\includegraphics[bb= 0 140 446 700,width=7cm]{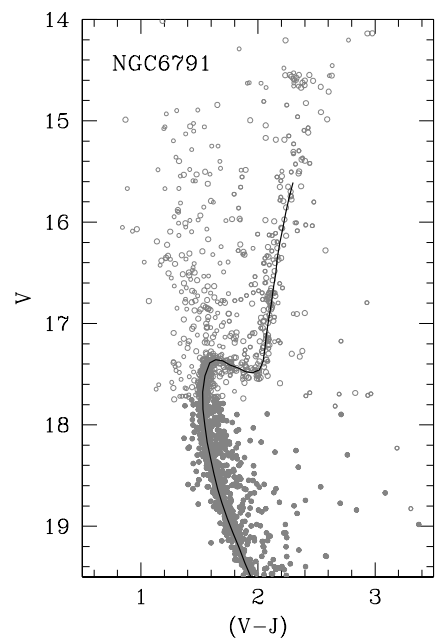}
\caption{The derived $V$-$(V-J)$ fiducial sequence of NGC$\,$6791 overlaying CFHT (filled circles) and 2MASS (open circles) photometry. }
\label{NGC6791_fiducial}
\end{figure}

\begin{figure}
\includegraphics[bb= 0 140 446 700, width=7cm]{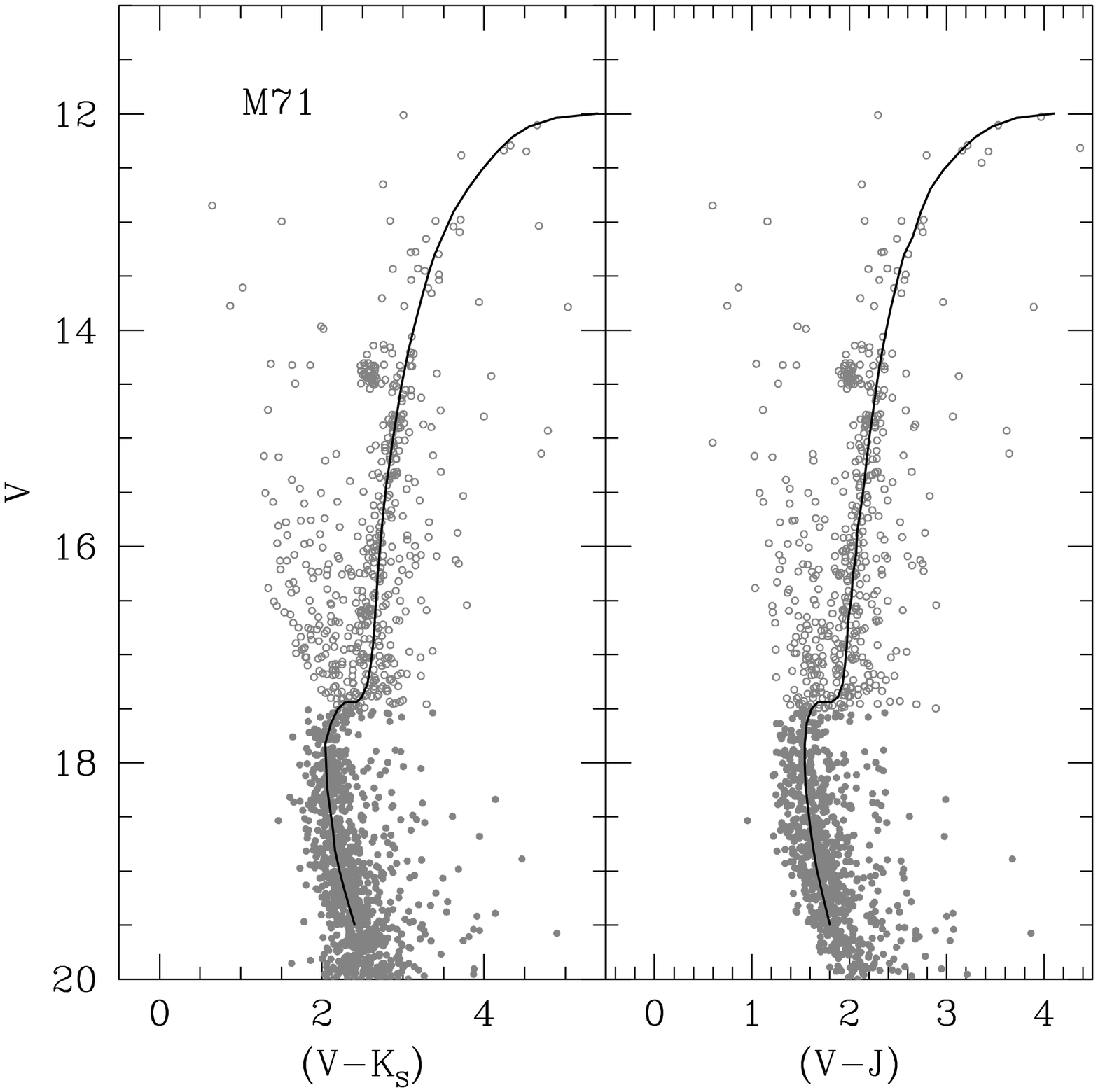}
\caption{The derived $V$-$(V-K_S)$ and $V$-$(V-J)$ fiducial sequences of M$\,$71 overlaying CFHT (filled circles) and 2MASS (open circles) photometry. }
\label{M71_fiducial}
\end{figure}

One can check the overall quality of our transformed magnitudes by comparing stars 
in common between our fields and the 2MASS catalog. One example of such a comparison is shown in Figure \ref{photometry} for NGC$\,$1851,  where the differences between the standard 
2MASS magnitudes and our final calibrated ones are plotted against both magnitude and colour.  
Reassuringly, the horizontal lines corresponding to zero difference appear to pass through the 
densest concentration of points in all plots. Moreover, there seem to be no strong systematic 
trends as a function of colour that would indicate the need for higher order colour terms in the 
photometric solutions.

\begin{figure}
\includegraphics[bb= 0 140 446 700,width=7cm]{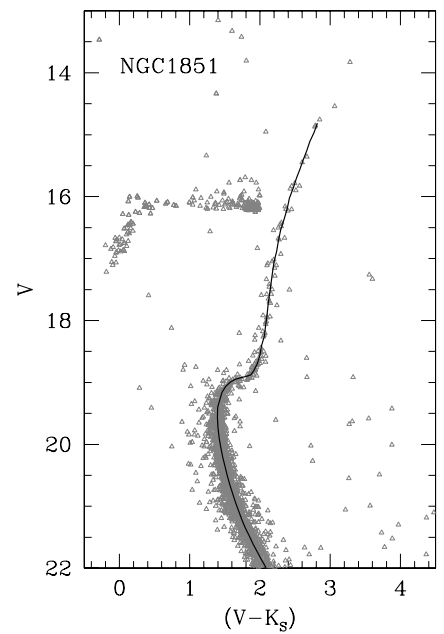}
\caption{The derived $V$-$(V-K_S)$ fiducial sequence of NGC$\,$1851 overlaying HAWK-I $K_S$ photometry.}
\label{NGC1851_fiducial}
\end{figure}

To produce the CMDs, our final photometry for each cluster was combined
with $V$-band data (see \citealt{Stetson2000}; \citealt{Stetson2005}).
Our CMD of NGC$\,$1851, which is from VLT observations, extends from near
the RGB tip to $\sim 3$ magnitudes below the MSTO.  For the remaining
six clusters, which were observed using the CFHT, we were able to obtain
observations only for magnitudes fainter than the base of the RGB.
The upper giant-branch stars were saturated in all of our frames, despite
observing the target clusters with the shortest exposures possible on
WIRCam,  This problem occurred mainly because the seeing was
0.3 to 0.5$^{\prime\prime}$ better than we had requested during the
nights of observation, which resulted in a more concentrated PSF.

Thus, in order to populate the RGBs of the clusters that were observed
using the CFHT, we queried the 2MASS catalog for all of the stars within
$30^{\prime}$ of the center of each cluster.  With a photometric
sensitivity of 10 sigma at $J$= 15.8 and $K_S$= 14.3 mag, 2MASS
observations extend at least to the base of the giant branch for each
of our target clusters. The 2MASS photometry for the selected giant-branch
stars in each system (\citealt{2MASS}) was then combined with $V$-band
photometry for the same stars (\citealt{Stetson2000}; \citealt{Stetson2005}).

Although it is ideal to have homogeneous observations for the entire
range in cluster magnitude, the zero-points of our WIRCam photometry
were set using 2MASS observations from the same catalog that was used
to identify the RGB stars. Therefore, in principle, one should expect
that there are no differences between these observations.  However, in
practice, the possibility that uncertainties in the zero-points may be
appreciable should be kept in mind when employing the resultant CMDs
and fiducial sequences.  A list of our observed clusters and our adopted reddening and 
distance moduli are given in Table 2.

\subsection{Fiducials}
\label{sec:Fiducials}

\begin{figure}[ht]
\includegraphics[bb= 0 140 446 786,width=7cm]{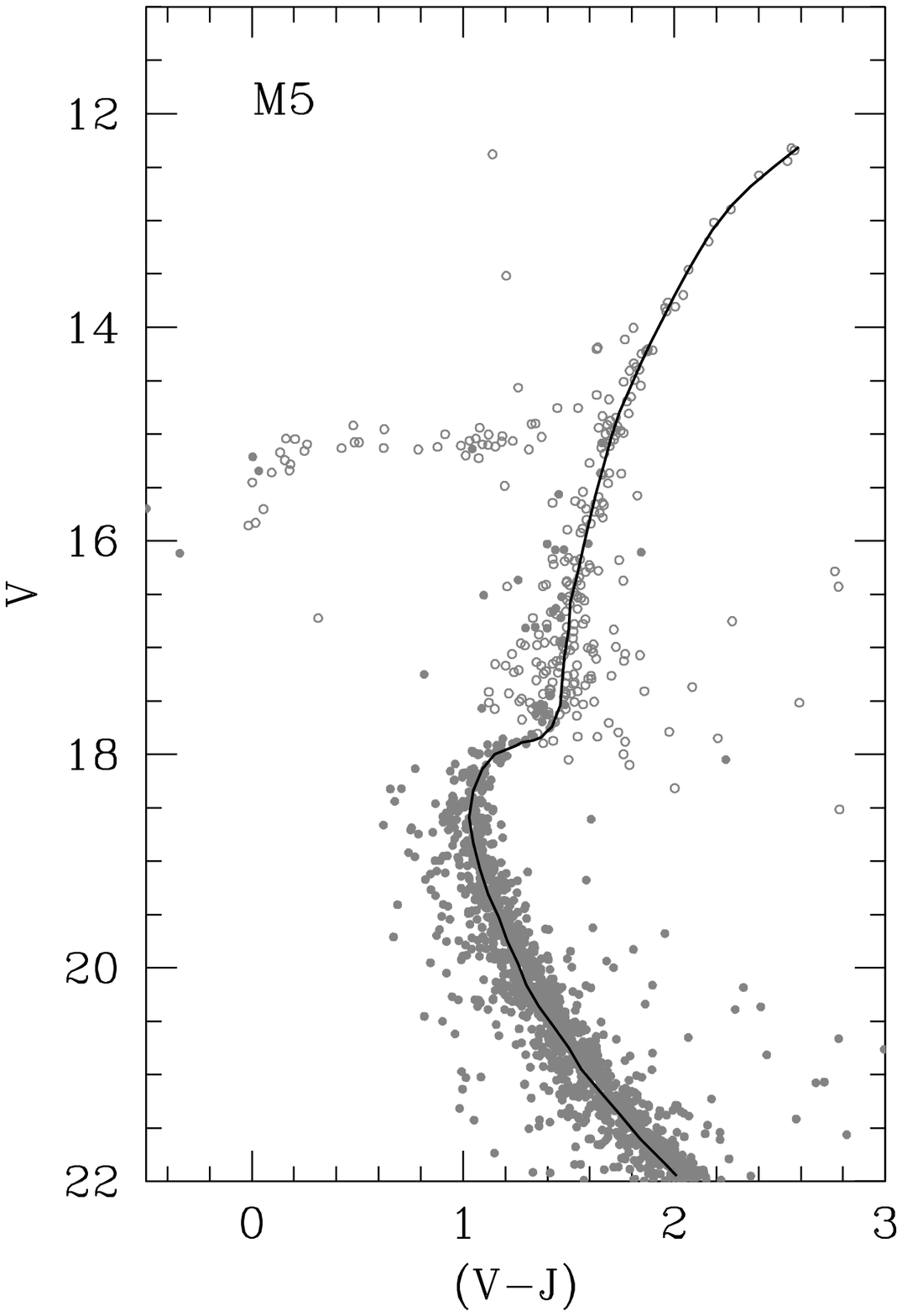}
\caption{The derived $V$-$(V-J)$ fiducial sequence of M$\,$5 overlaying CFHT (filled circles) and 2MASS (open circles) photometry.}
\label{M5_fiducial}
\end{figure}

Fiducial sequences are ridge lines of the stellar loci in
colour-magnitude space.  The definition of these fiducials from
cluster photometry is often based on visual inspection of the CMD,
since automated scripts typically give poor results in regions of low
star counts and where the magnitude varies weakly with colour, such as
the subgiant branch (SGB).  Moreover, contamination from field, AGB,
binary, and horizontal-branch (HB) stars can significantly skew the
the computed locus.  For these reasons, we have derived all fiducials by
dividing the magnitude axis into small bins (typically $\sim 0.15$ mag
wide, or smaller in regions of nearly constant magnitude) and then
calculating the median colour of those stars which we judge to belong
to the cluster.

\begin{figure}
\includegraphics[bb= 0 140 446 700, width=7cm]{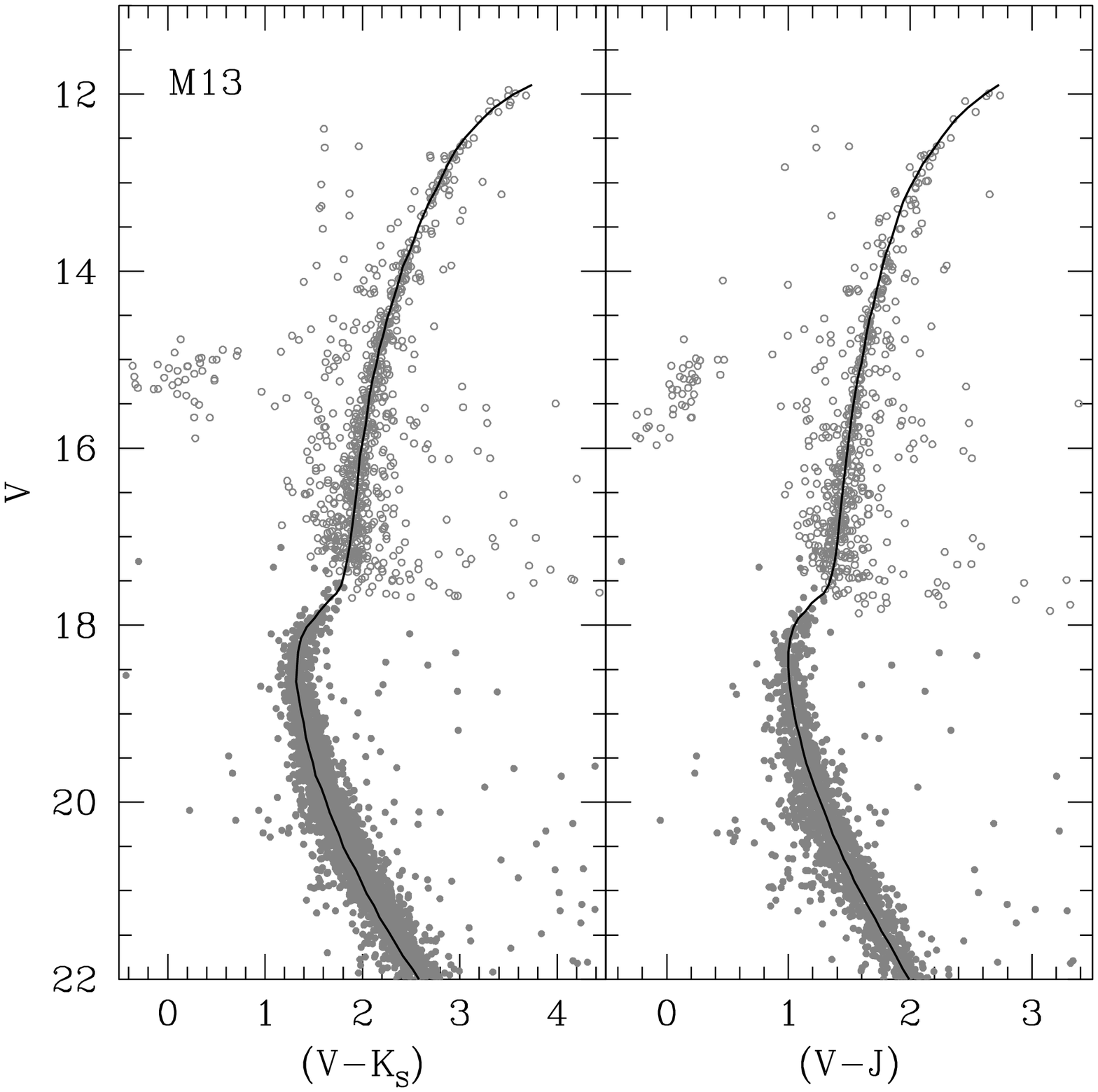}
\caption{ The derived $V$-$(V-K_S)$ and $V$-$(V-J)$ fiducial sequences of M$\,$13 overlaying CFHT (filled circles) and 2MASS (open circles) photometry. }
\label{M13_fiducial}
\end{figure}

\begin{figure}
\includegraphics[bb= 0 140 446 700, width=7cm]{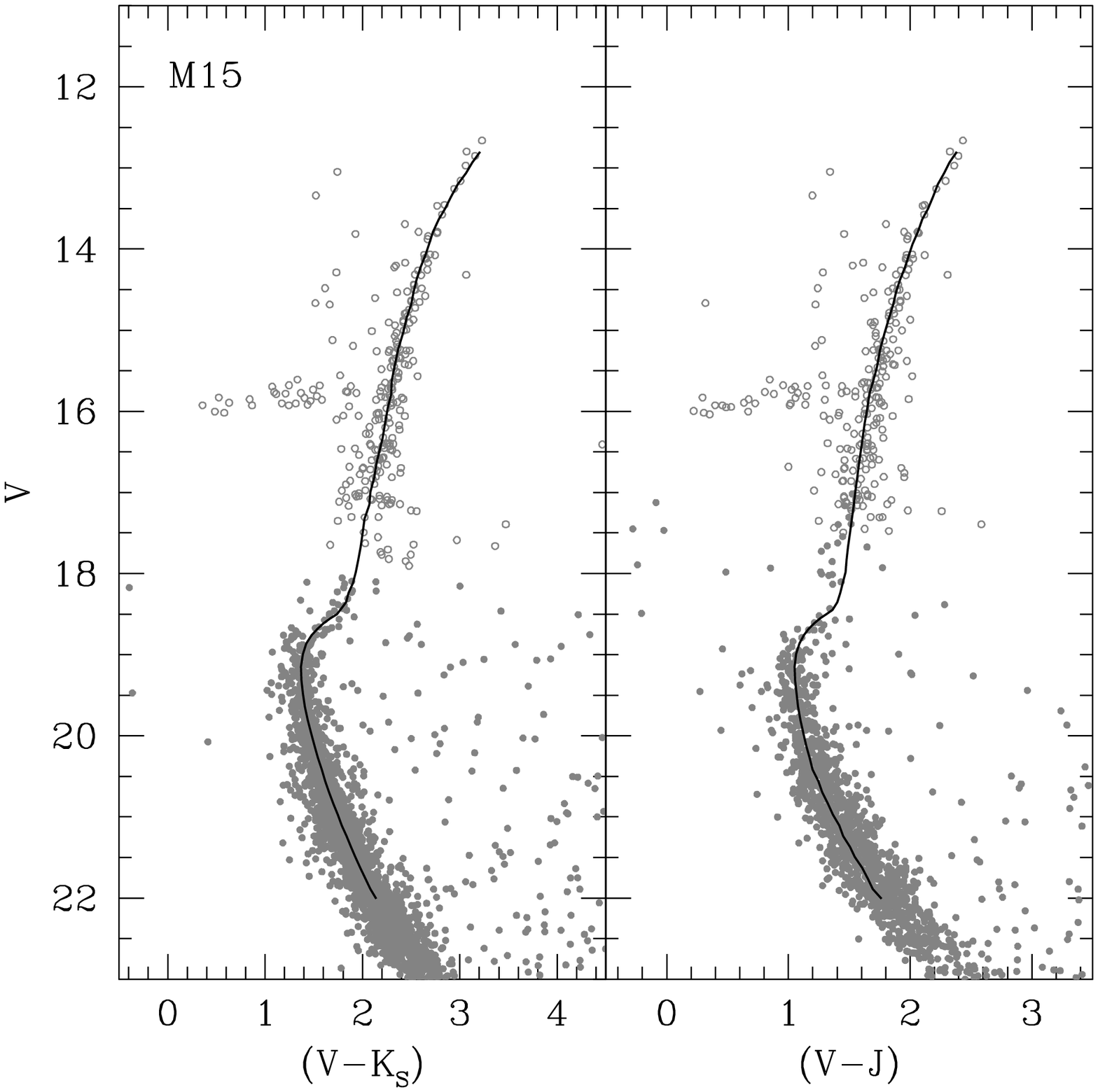}
\caption{ The derived $V$-$(V-K_S)$ and $V$-$(V-J)$ fiducial sequences of M$\,$15 overlaying CFHT (filled circles) and 2MASS (open circles) photometry.}
\label{M15_fiducial}
\end{figure}

\begin{figure}
\includegraphics[bb= 0 140 446 700, width=7cm]{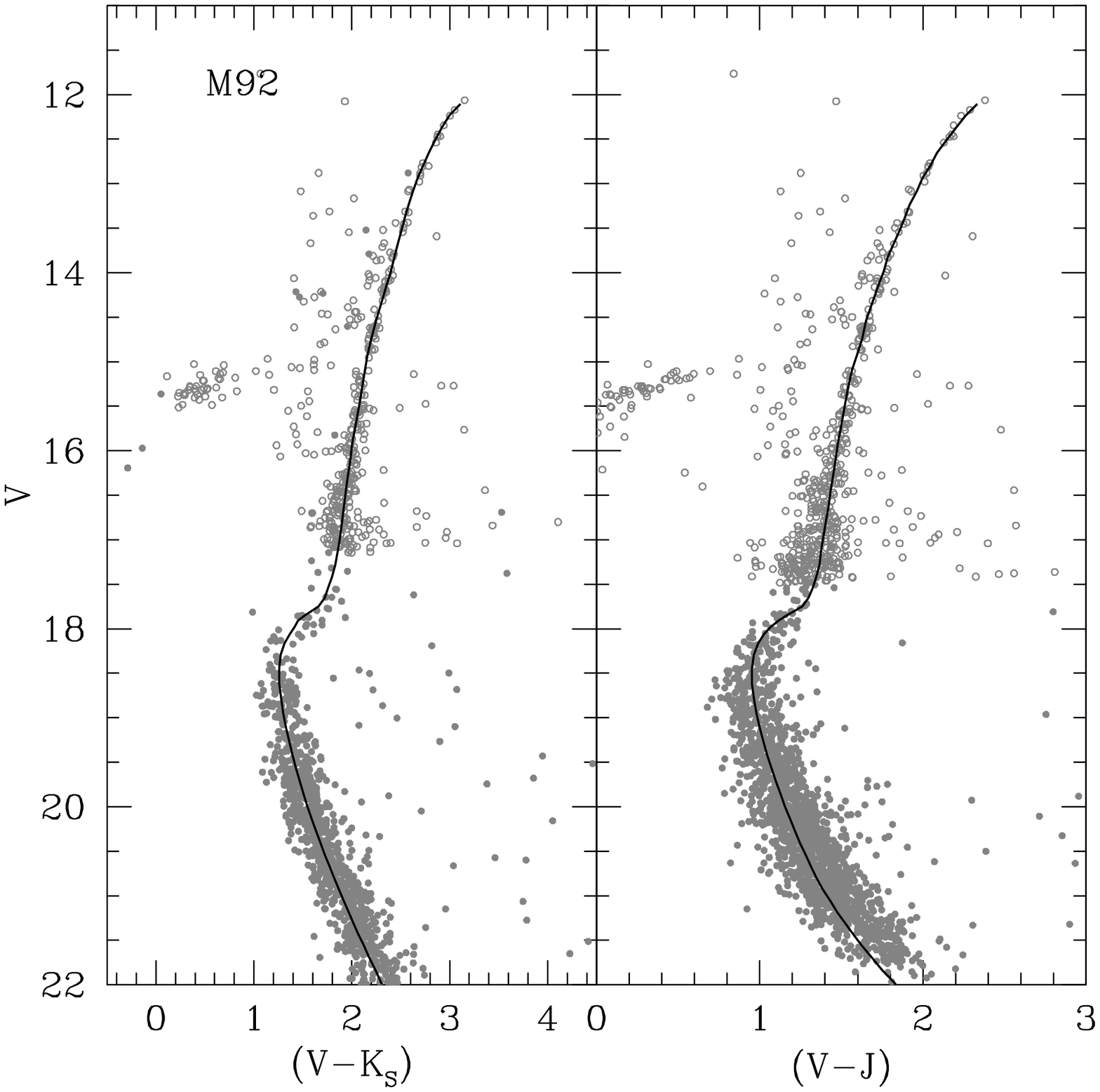}
\caption{The derived $V$-$(V-K_S)$ and $V$-$(V-J)$ fiducial sequences of M$\,$92 overlaying CFHT (filled circles) and 2MASS (open circles) photometry. }
\label{M92_fiducial}
\end{figure}

We present our observed CMDs in Figures
\ref{NGC6791_fiducial}--\ref{M92_fiducial} with 2MASS, CFHT/WIRCam,
and VLT/HAWK-I photometry plotted as open circles, filled circles,
and open triangles, respectively.  Overlaid on the colour-magnitude
diagrams are the derived fiducial sequences spanning the MS, SGB,
and RGB (see Tables 2--9).   In the case of M$\,$13, M$\,$92, and
M$\,$15, there were insufficient points to define the fiducial for
the subgiant branch; consequently, the latest Victoria-Regina Models
 by VandenBerg et al. (in preparation)  of the same metallicity of each cluster were used to define
the transition from the MS to the lower RGB in these cases (such
points are marked with an asterisk in the tables).

\begin{table*}
 \hspace{0.033\textwidth}
 \centering
\begin{tabular}{l c c c c c c} \hline \hline
Cluster & Type & $\alpha$ &$\delta$ &  [Fe/H] & (m-M)$_V$&E($B-V$)  \\
  \hline
M$\,$15 & globular &   21:29:58  & +12:10:01&$-$2.4 & 15.29  & 0.108 \\
M$\,$92 & globular & 17:17:07 & +43:08:12 & $-$2.4 & 14.62 &0.023 \\
M$\,$13& globular  &16:41:41& +36:27:37 & $-$1.60 & 14.40 & 0.016\\
M$\,$5& globular  &15:18:36& +02:05:00 & $-$1.40 & 14.45 & 0.038\\
NGC 1851 &globular & 05:14:06 &-$~$40:02:50 & $-$1.40 & 15.50& 0.034\\
M$\,$71 &globular & 19:53:46 & +18:46:42 & $-$0.80 & 13.78 & 0.220\\
NGC$\,$6791& open & 19:20:53 & +37:46:30 & +0.3&13.57 &  0.150\\

\hline
 \end{tabular}
\caption[Properties of the Galactic star clusters in our survey.]{Properties of the Galactic star clusters in our survey.}
\label{cluster_params}
\end{table*}

\section{Testing the Colour-$\Teff$ Relations using Observations of Field
Subdwarfs and Star Clusters}
\label{sec:ctrelations}

Transforming isochrones from the theoretical $\log\Teff$--M$_{bol}$ plane
to an observed CMD is accomplished through the use of colour--$\Teff$
relations to link the fundamental stellar parameters to photometric indices.
In this section, we investigate two recently developed colour transformations.
The first of these is an empirical relation developed by \cite{Casa2010}
(hereafter referred to as CRMBA) who have applied the IRFM to a large sample
of dwarf and subgiant stars of varying metallicity.  They present their
results as a set of polynomials which relate many photometric indices to
$\Teff$ and [Fe/H].  This empirical colour--$\Teff$ relation has the
advantage of being largely model independent, but it has the limitation of
being applicable only to dwarf and SGB stars.
 
The second relation was derived by one of us (L.C.) from synthetic
spectra based on the latest MARCS model atmospheres (\citealt{Gust2008})
using the procedures described 
in VCS10, with the 2MASS filter transmission curves, absolute calibration and 
zero-points reported in \citet{casa2006}.  The MARCS models consist of
plane-parallel (for dwarf and SGB stars) and spherical (for giant stars)
line-blanketed, flux-constant stellar atmospheres for large ranges in
effective temperature, surface gravity, and chemical composition.  By
convolving the spectra derived from these atmospheres with the appropriate
filter transmission functions, large tables were produced that provide
the synthetic magnitudes as functions of [Fe/H], $\Teff$, and $\log g$.
Theoretical relations such as these have the advantage over the CRMBA
relation that the desired photometric indices for any given star or
stellar model can be obtained simply by interpolating it in the grid of
synthetic colours for its [Fe/H], $\log\,g$, and $\Teff$ values.
However, their accuracy relies heavily on whether or not the synthetic
spectra are able to reproduce the observed spectra of stars, and on how
well the filter transmission functions and zero-points are defined.  It should also be noted that
we apply the slight modifications to the 2MASS model calibration noted in the recent CRMBA paper. These very slightly affect the $JHK_S$ model colours ($J$ $\rightarrow$ $J$ - 0.017; $H$  $\rightarrow$ $H$ + 0.016; $K_S$  $\rightarrow$$K_S$ + 0.003) and indeed these improve the fits to the data.

In the following analysis we transform the latest Victoria isochrones
(VandenBerg et al. in preparation) using the MARCS and CRMBA
colour-$\Teff$ relations and test how well they are able to reproduce
the observed $VJK_S$ colours of field subdwarfs and our cluster
photometry.   
Whenever possible, distances derived from {\it Hipparcos}
parallaxes, along with the most up-to-date estimates of the cluster
metallicities, have been assumed.  However this study is much less
concerned with the absolute fit of the isochrones to the observed CMDs
than with the extent to which a {\it consistent} interpretation of the
same cluster can be found on different colour planes.  Since any
inaccuracies in distance moduli or [Fe/H] should be evident in
{\it all} colour planes, they should not effect our overall conclusions
(though the possibility that chemical abundance anomalies may be present,
and that they might affect the fluxes in some filter bandpasses more
than others should be kept in mind).  This does, of course, have the
caveat that the observations must be free from systematic errors and
that the extinction in all colour bands is accurately determined.

\begin{figure}
\includegraphics[bb= 0 140 446 700, width=7cm]{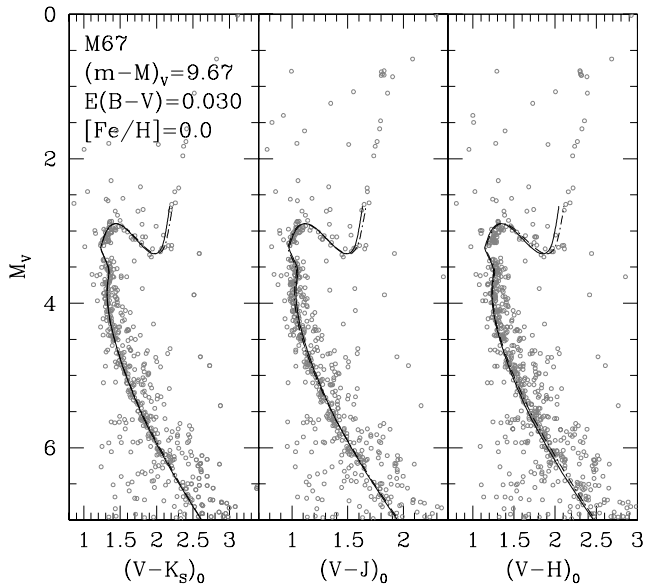}
\caption{Comparison of an 3.7 Gyr isochrone for [Fe/H] $= 0.0$ with
2MASS/Stetson $V-K_S$, $V-J$, and $V-H$ observations of M$\,$67.  The
solid and dot-dashed curves assume the MARCS and \citet{Casa2010} colour
transformations, respectively. }
\label{M67_marcs}
\end{figure}

As previously mentioned, this paper provides an extension of the
BV$(RI)_C$ study by VCS10 to the near-IR.  Consequently, we have tried
to ensure that the analysis of the colour transformations for a given
cluster is consistent across optical and infrared colours by employing
identical isochrones (i.e., for the same age and metallicity) and by 
adopting the same values of the reddening and distance modulus (in the case
of clusters which are in common with VCS10: M$\,$67, NGC$\,$6791, NGC$\,$1851, M$\,$5, M$\,$92).   Additionally, the $V$-band
photometry of the clusters presented in the following sections is from
the same catalogues (\citealt{Stetson2000}; \citealt{Stetson2005}) that
were used for the comparisons presented by VCS10.

The extinction coefficients adopted in this work are  E($V-J$)/E($B-V$)=2.25, E($V-K_S$)/E($B-V$)=2.76 and E($V-H$)/E($B-V$)=2.55 (\cite{McCall2004}).

\subsection{M$\,$67 ([Fe/H] $\approx 0.0$)}
\label{sec:M67}

M$\,$67 is a particularly well studied, relatively nearby open cluster,
with a metallicity that is very close to solar, according to the results of high-resolution spectroscopy (\citealt{Tautv2000}; \citealt{Randich2006}).  See VCS10 for a summary 
of its basic properties, together with supporting references.  Because
of its proximity, 2MASS $JHK_S$ observations of M$\,$67 reach well below
the MSTO; consequently, the CMDs shown in Figure~\ref{M67_marcs} could
be produced simply by combining M$\,$67 data from 2MASS (\citealt{2MASS})
with the latest reduction of the $V$-band photometry discussed by
\citet{Stetson2005}.  The isochrone in this figure is that for an age
of 3.7 Gyr from \cite{Michaud2004}, and just as VCS10 found from their
comparisons of the same isochrone with $BV(RI)_C$ photometry, both the
MARCS and CRMBA transformations to the $V-J$ and $V-K_S$ colour planes 
enable the models to match the observed photometry exceptionally well
over the entire range of luminosity that has been plotted.  Indeed, the
differences between the two colour--$\Teff$ relations are barely
discernible.  (Only in the case of M$\,$67 have we compared theory and
observations on a $[(V-H)_0,\,M_V]$-diagram.  Although the isochrone
fits the data quite well, there is a slight, apparently nearly constant 
offset between the two, which may be due to a small error in our assumed
value of $E(V-H)$ or in the zero-point of the $H$ magnitudes derived 
from the MARCS model atmospheres.)

\begin{figure}
\includegraphics[bb= 0 140 446 700, width=7cm]{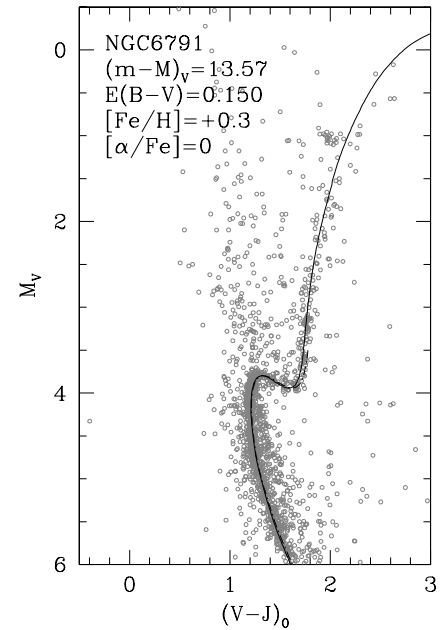}
\caption{Comparison of an 8.0 Gyr isochrone for [Fe/H] $= +0.3$ and $Y=0.30$
with our $V-J$ observations of NGC$\,$6791.  The solid and dot-dashed curves
assume the MARCS and \citet{Casa2010} colour transformations, respectively.}
\label{NGC6791_marcs}
\end{figure}

\subsection{NGC$\,$6791 ([Fe/H] $\approx +$0.3)}
\label{sec:NGC6791}

Being the most metal rich cluster in our sample (and only open cluster
for which we have collected $JK_S$ observations), NGC$\,$6791 provides an
important test of the colour--$\Teff$ relations applicable to super-metal-rich
stars.  Here we have adopted the recent [Fe/H] estimate from \cite{Boesgaard2009}. 
This is consistent with \citet{Brogaard2010} who derived
[Fe/H] = 0.29 $\pm$ 0.03 (random) $\pm$ 0.07 (systematic) in their
spectroscopic analysis.  \cite{Anthony2007} have noted
that the \cite{Carraro2006} determination of [Fe/H] = +0.38 for
NGC 6791 should be reduced to $\sim$ +0.28 (hence in very good agreement
with the Brogaard et al. result) if those authors had adopted E(B-V) =
0.15 in their analysis, instead of 0.09.

Because of its large distance and low Galactic latitude,
reddening estimates of NGC$\,$6791 vary considerably in the literature.
As discussed by \cite{Chaboyer1999}, the derived reddenings for
NGC$\,$6791 span the range $0.09\leq E(B-V) \leq 0.26$.  \cite{Kuch95}
 have used subdwarf-B stars to provide a tight
constraint on the reddening, finding $E(B-V)= 0.17\pm0.01$.  This
agrees well with the \cite{Schlegel98} estimate of $E(B-V)= 0.155$
mag, which is also favoured by the comparison of the NGC$\,$6791 CMD
with that for solar neighborhood stars from {\it Hipparcos} data
(see \citealt{Sandage2003}).  Additionally, \cite{Brogaard2010} also derived E(B-V) = 0.160 $\pm$ 0.025, which adds to the support of our adopted reddening of 0.15.

In consistency with VCS10, we adopt a distance
modulus of $(m-M)_V= 13.57$ which was found through a fit of the cluster CMD to local
field dwarfs having metal abundances in the range $+0.15\leq$ [Fe/H]
$\leq +0.45$ and $\sigma(M_V) \leq 0.15$ (based on {\it Hipparcos}
parallaxes).  Figure \ref{NGC6791_marcs} shows our $[(V-J)_0,\,M_V]$
CMD for dwarf and SGB stars in NGC$\,$6791 is well reproduced by either
the MARCS- or CRMBA-transformed Victoria isochrone on the assumption of the 
aforementioned reddening and distance.  (Only along the upper RGB does
the isochrone seem to deviate to the blue of the observations, which is consistent
with the possibility that the MARCS atmospheres have insufficient
blanketing in cool super-metal-rich stars.)  Thus, there is excellent
consistency with the isochrone fits reported by VCS10 to the
$[(B-V)_0,\,M_V]$- and  $[(V-I)_0,\,M_V]$-diagrams of NGC$\,$6791.
Since $E(V-J) = 2.25\,E(B-V)$ (e.g., \citealt{McCall2004}), this consistency
provides a further (strong) argument that the foreground reddening must,
indeed, be quite close to $E(B-V) = 0.15$.  This example thus provides
an instructive example of how valuable it is to have both optical and
near-IR photometry to constrain the reddening in the case of highly
reddened systems.

\subsection{M$\,$71 ([Fe/H] $\approx -$0.75)}
\label{sec:M71}

\begin{figure}
\includegraphics[bb= 0 140 446 700, width=7cm]{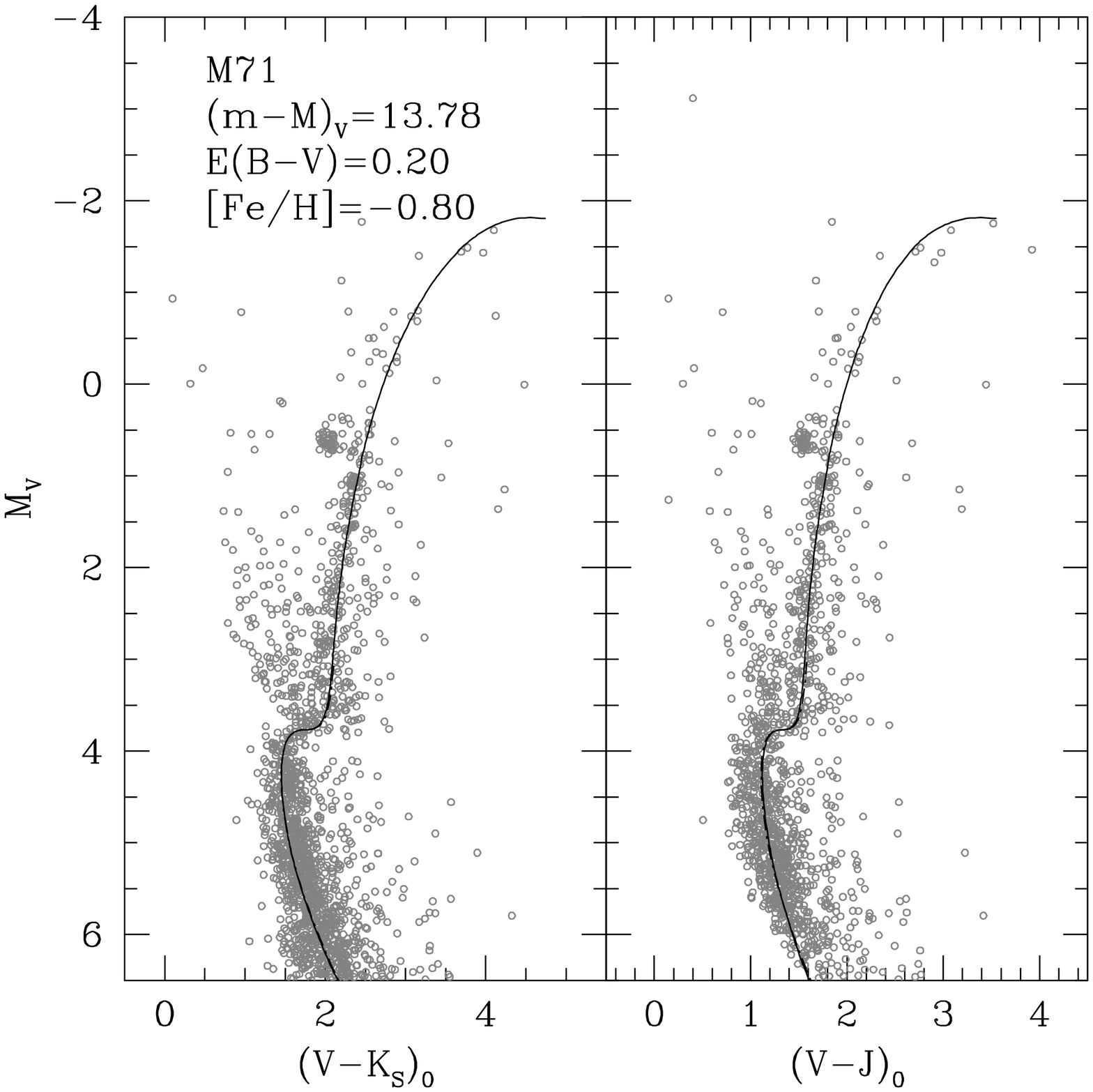}
\caption{Comparison of an 11 Gyr isochrone for [Fe/H] $= -0.75$, with
our observations of M$\,$71. The solid and dot-dashed curves assume the
MARCS and \citet{Casa2010} colour transformations, respectively.  }
\label{M71_marcs}
\end{figure}

M$\,$71 represents the metal-rich end of our sample of globular
clusters (GCs). We adopt an [Fe/H]=$-$0.80 as derived by \citet{Boesgarrd2005}.
 It is quite obvious from the CMD in
Figure \ref{M71_fiducial} that this cluster has a low Galactic 
latitude, as contamination from field stars is still significant
even when only those stars contained within an annulus of $\sim$7 arcmin around the
cluster center are plotted.  Furthermore, the \cite{Schlegel98}
dust maps indicate that the reddening is differential across the cluster,
which contributes to the spread in colour at any given magnitude,
thereby hindering the definition of tight photometric sequences
for this cluster in any photometric band. 

As far as the cluster distance is concerned, we have adopted a
distance modulus that leads to the most consistent interpretation
between M$\,$71 (in this paper), and that of 47 Tuc (in VCS10), which
is known to have very close to the same metallicity.  If the latter has
$(m-M)_V=13.40$, as assumed by VCS10, and if the difference in magnitude
of the HBs of 47 Tuc and M$\,$71 is as given by \cite{H992},
then a value of $(m-M)_V = 13.78$ is obtained for M$\,$71. 

Because of its low Galactic latitude (b =$-$4.6$^\circ$), M71 poses a challenge for reddening determinations: estimates in the literature range from E($B-V$)= 0.21 to 0.32 (e.g. Kron \& Guetter 1976;  Harris 1996). 

Figure \ref{M71_marcs} shows the isochrones compared with our observations.  
The observations are matched well by isochrones on the $[(V-K_S)_0,\,M_V]$-plane using either the MARCS or CRMBA colour transformations.  In the case of the $V-J$ observations, the isochrones match the base of the RGB well, but drift to the
blue side of the lower MS, and to the red of the upper MS.  However, the discrepancies at
the faintest magnitudes may have an observational origin given
that they arise close to the photometric limit of $J$. It should be noted that the reddening adopted here, $E(B-V)=0.20$ was chosen to obtain a fit to the $B-V$ photometry. Thus, the models provide consistent interpretations across the $B-V$, $V-J$ and $V-K$ colour-planes.

\begin{figure}
\includegraphics[bb= 0 140 446 700, width=7cm]{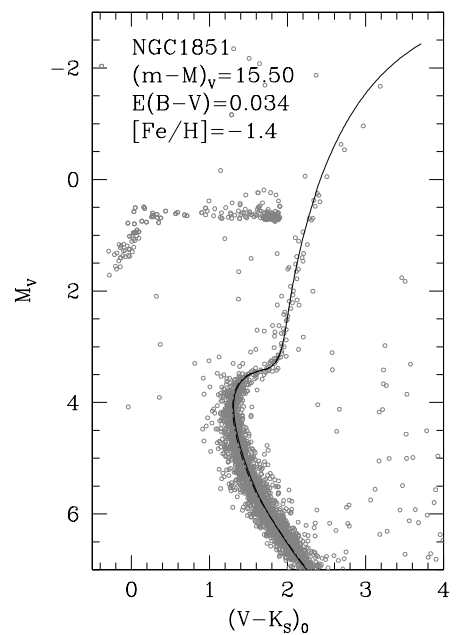}
\caption{Comparison of an 11 Gyr isochrone for [Fe/H] $= -1.4$,
[$\alpha$/Fe] $\approx 0.4$, and $Y=0.25$ with our $V-K_S$ observations of
NGC$\,$1851. The solid and dot-dashed curves assume the MARCS and
\citet{Casa2010} colour transformations, respectively. }
\label{NGC1851_marcs}
\end{figure}

\subsection{M$\,$5 and NGC$\,$1851 ([Fe/H] $\approx -$1.4)}
\label{sec:M5}

As discussed quite extensively by VCS10, when Schlegel et al.~(1998) reddenings are applied to the observed stars in M$\,$5 and NGC$\,$1851 and then their CMDs are shifted in the vertical direction so that the red HB populations of the two clusters have the same luminosities, one finds that the main-sequence fiducials of both clusters superimpose one another nearly perfectly, suggesting a common metallicity.  This conflicts with the results of most spectroscopic studies, which  indicate that NGC$\,$1851 is $\approx 0.2$ dex more metal-rich than M$\,$5 (e.g., Carretta et al.~2009; \citet{KraftIvans}).  If this is correct, the latter should have a slightly brighter HB than the former, and therefore a slightly brighter MS at a given color --- but this would be inconsistent with the assumed difference in [Fe/H] (unless $Y$ or the heavy-element mixture also varies).  However, a resolution of this issue is not important for the present investigation, which seeks to determine if consistent interpretations of the photometry across all colour planes can be found.  Following VCS10, we adopt [Fe/H]=$-$1.4 for both M$\,$5 and NGC$\,$1851 which is the metallicity which tends to be favoured by stellar models such as \citet{V00} (see VCS10 for a detailed description on this choice of metallicity).

\begin{figure}
\includegraphics[bb= 0 140 446 700, width=7cm]{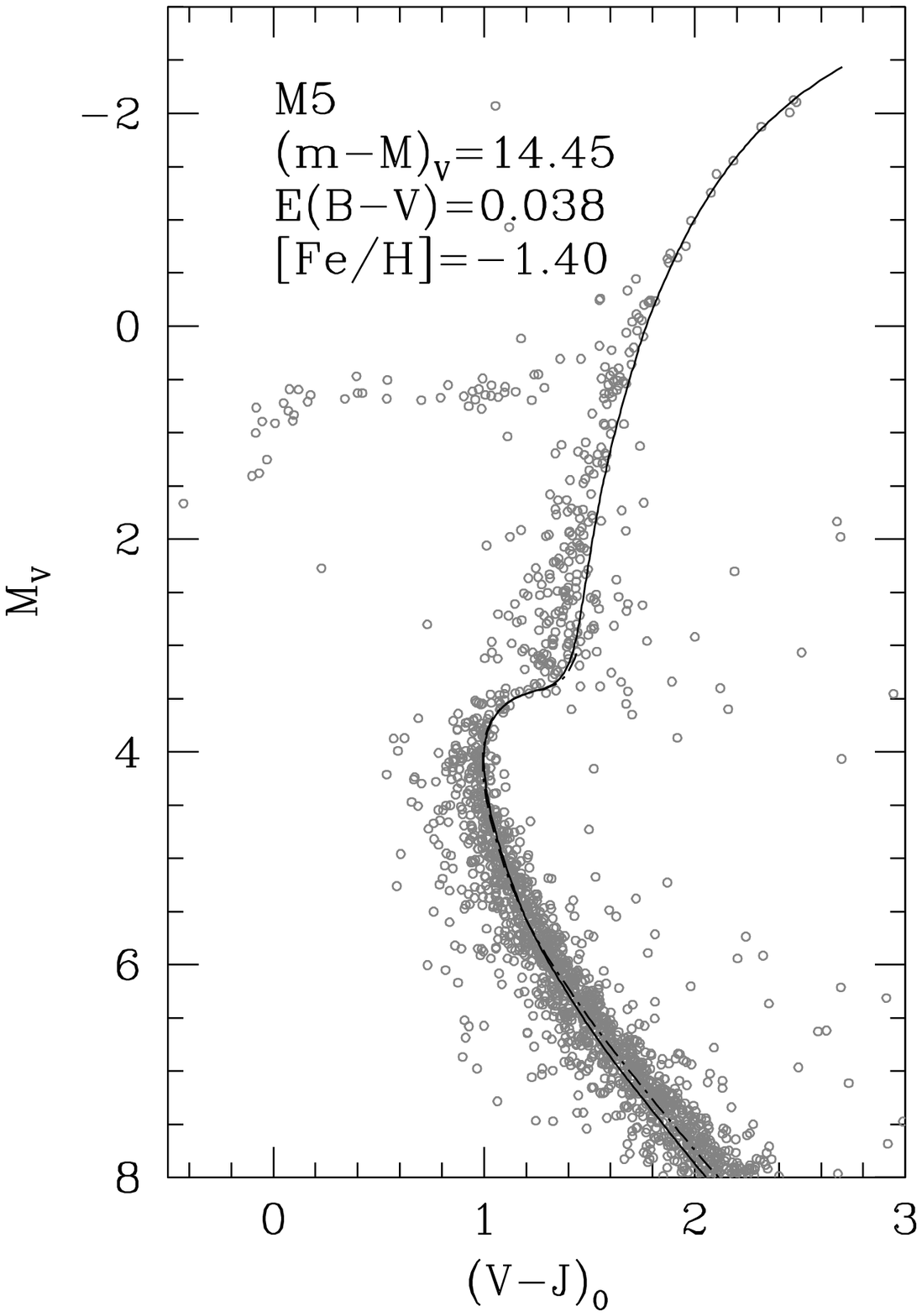}
\caption{As in the previous figure, except that the (same) isochrone
is compared with our $V-J$ observations of M$\,$5. }
\label{M5_marcs}
\end{figure}

In Figure~\ref{NGC1851_marcs}, we overlay the same 11 Gyr isochrone used
by VCS10 in their study of $BV(RI)_C$ photometry, onto the 
$[(V-K_S)_0,\,M_V]$-diagram of NGC $\,$1851. For consistency with VCS10,
we have adopted the same reddening and distance modulus; specifically,
$E(B-V)=0.034$ (\citealt{Schlegel98}) and $(m-M)_V = 15.50$, based on
observed and predicted HB luminosities (\citealt{Vetal00}) as well as
MS-fits of cluster fiducials to the CRMBA-transformed isochrones (and hence to the subdwarf standards for which these relations are defined;  VCS10).  We find that isochrones using the MARCS
colour--$\Teff$ relations, or the CRMBA transformations for dwarf/SGB
stars, are able to reproduce our $V-K_S,\,V$ photometry of NGC$\,$1851
rather well, except for the upper RGB (where there are few stars and where
the stellar images are approaching the saturation limit).  These results
are fully consistent with those found by VCS10 when they analyzed
$V-I,\,V$ data.  However, as they reported, the giant branch of the
isochrone on the $[(B-V)_0,\,M_V]$-diagram is significantly bluer than
the observed RGB.  Thus, only the MS stars can be consistently fitted on
all of the colour planes that have been considered.

In the case of M$\,$5, we have assumed that $E(B-V)=0.038$ and $(m-M)_V
= 14.45$ (see VCS10), which results in the comparison of the MARCS-
and CRMBA-transformed isochrones with the $V-J$ observations shown in
Figure \ref{M5_marcs}.  (Note that the isochrones are identical to those
fitted to NGC$\,$1851 photometry.)   Given the differences in the 
relative positions of the RGBs of M$\,$5 and NGC$\,$1851 noted above,
it is not surprising to find that, since the MARCS models are able to
reproduce the giant branch of NGC$\,$1851 in $V-K_S$ and $V-I$, they
lie to the red of the observed RGB of M$\,$5 in $V-J$ and $V-I$ (see
VCS10, who also show that the predicted giant branch matches the $B-V$
colours of M$\,$5 giants quite well).  As far as the main sequence of
of M$\,$5 on the $[(V-J)_0,\,M_V]$-diagram is concerned, the isochrones
lie slightly blueward of the lower MS, and to the red of the upper MS, as seen in M71.  As discussed
below, this is also seen in the case of M$\,$15 and M$\,$92, and
perhaps indicates that there is a small problem with the transformations
to $V-J$.

\subsection{M$\,$13 ([Fe/H]$\approx -$1.6)}
\label{sec:M13}

\begin{figure}
\includegraphics[bb= 0 140 446 700, width=7cm]{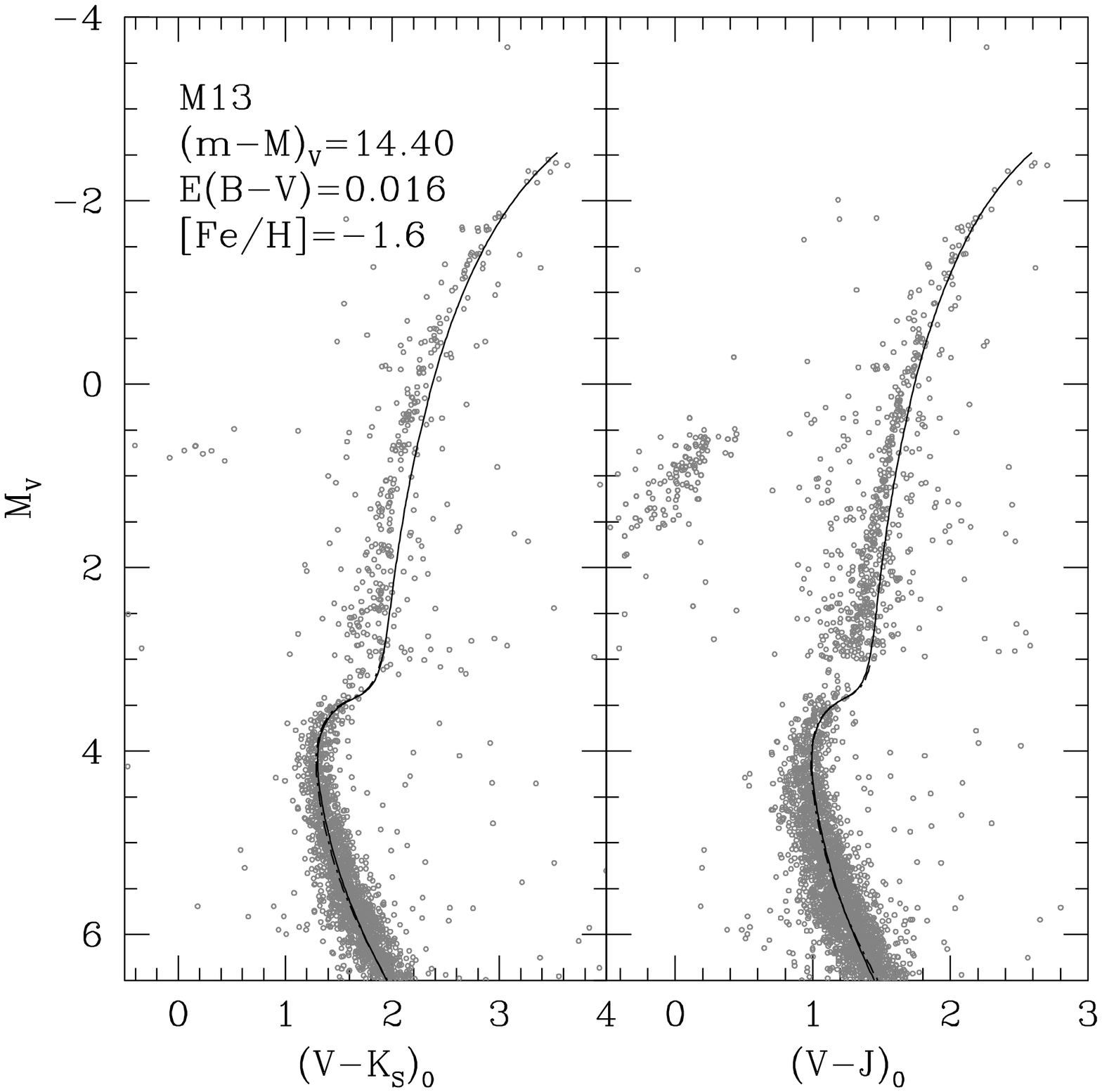}
\caption{Comparison of a 12 Gyr isochrone for [Fe/H] $= -1.60$ with our
observations of M$\,$13. The solid and dot-dashed curves assume the MARCS
and \citet{Casa2010} colour transformations, respectively. }
\label{M13_marcs}
\end{figure}

For our comparisons with M$\,$13, we have selected isochrones with
[Fe/H] $= -1.6$ based on the work of \citet{KraftIvans}.  This is within
0.05 dex of the long favoured \citet{zw84} value of $-1.65$ and within
0.02 dex of the recent \citet{Carretta2009} value of $-1.58$.  We have
also adopted $(m-M)_V =14.40$ which is very close to the value obtained by
\cite{Grun1998} from a fit of Stromgren photometry to metal-poor
subdwarfs with Hipparcos parallaxes, who also assumed [Fe/H] = -1.6. Additionally, \citet{Paltrinieri1998} derived $\delta$(V) between M$\,$3 and M$\,$13 of 0.64 mag
- so if M$\,$3 has 15.00 (as adopted in VCS10), then 14.40 is implied for M$\,$13 from the
Paltrinieri et al result (if the 0.01 mag difference in reddening is taken into account).  We adopt $E(B-V)=0.016$ from \cite{Schlegel98}.

In figure \ref{M13_marcs}, we compare a 12 Gyr isochrone with our $VJ$ and $VK_S$ photometry,
finding that (for both colours) the isochrones provide a good match
to the observed MS, but not to the observed RGB, which is significantly
bluer at a fixed magnitude than the predicted colours.  This is also
found when fitting $VI$ data (not shown here); i.e., when the same
isochrones are overlaid onto $V-I,\,V$ photometry, the main sequence
is well matched but the predicted giant branch is offset to the red of 
the observed RGB.  Interestingly, no such problem is found fitting the
$VI$ (or $BV$) photometry of M$\,$3 (see VCS10), which is thought to
have close to the same metal abundance as M$\,$13.

\subsection{M$\,$15 and M$\,$92 ([Fe/H] $\approx-$2.4)}
\label{M92_M15}

\begin{figure}
\includegraphics[bb= 0 140 446 700, width=7cm]{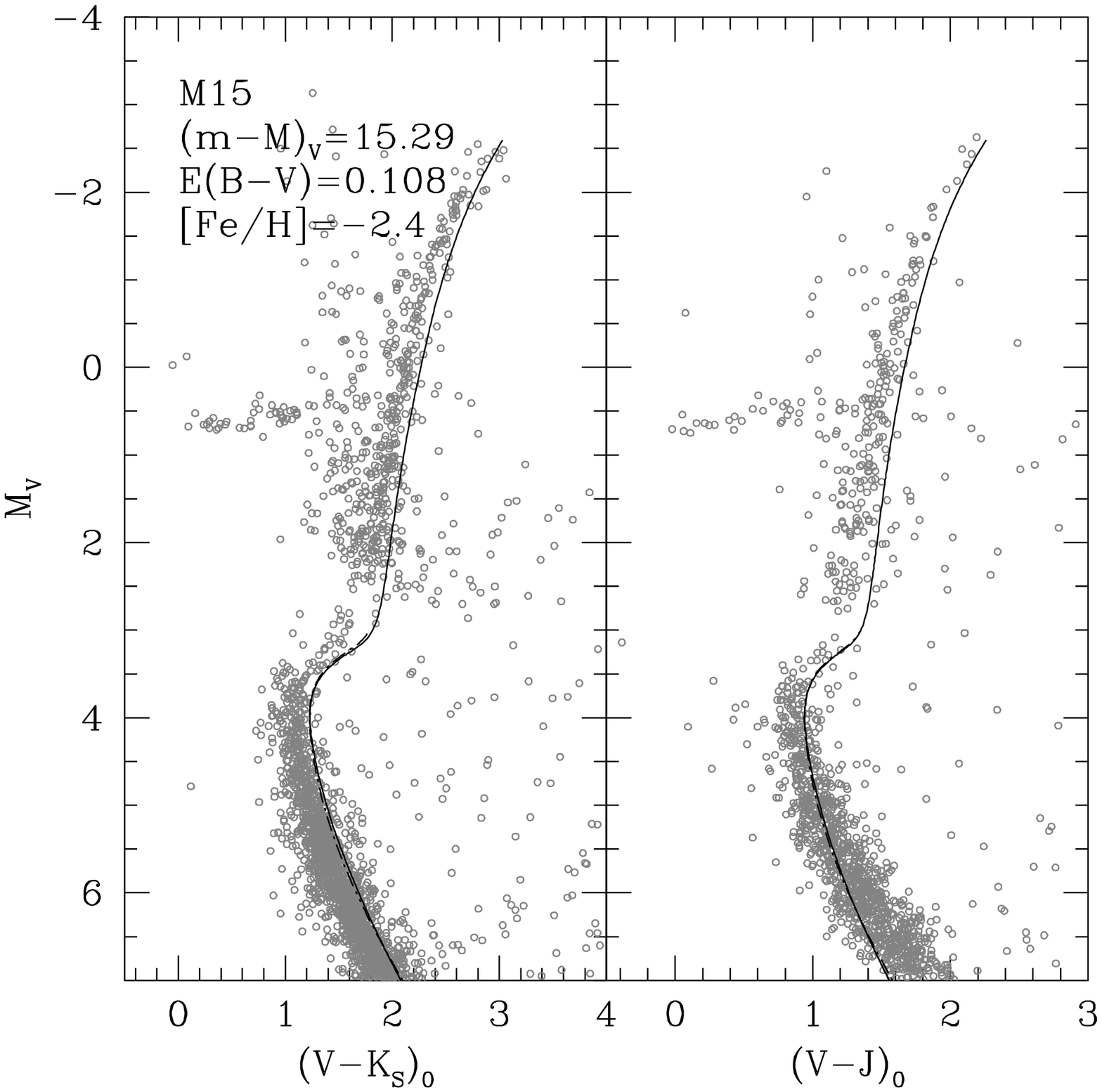}
\caption{Comparison of a 13.5 Gyr isochrone for [Fe/H] $= -2.40$ with
our observations of M$\,$15. The solid and dot-dashed curves assume the
MARCS and \citet{Casa2010} colour transformations, respectively. }
\label{M15_marcs}
\end{figure}

M$\,$15 and M$\,$92 constitute the most metal poor clusters in our
sample.  For both clusters, a metallicity near [Fe/H] $=-2.4$ is 
consistent with the latest results from high-resolution spectroscopy
(\citealt{KraftIvans}; \citealt{Carretta2009}).  As regards their distances,
we have adopted $(m-M)_V = 14.62$ for M$\,$92 (see \citet{VandenBerg2002}),
and $(m-M)_V = 15.29$ for M$\,$15 given that \citet{D93}
have found that the main sequences of both clusters will superimpose
one another if a vertical shift of 0.67 mag is applied to the M$\,$92
CMD once their turnoff colours are matched.  According to \cite{Schlegel98}, 
the reddenings of M$\,$92 and M$\,$15 are $E(B-V) = 0.023$
and 0.108 mag, respectively.

\begin{figure}
\includegraphics[bb= 0 140 446 700, width=7cm]{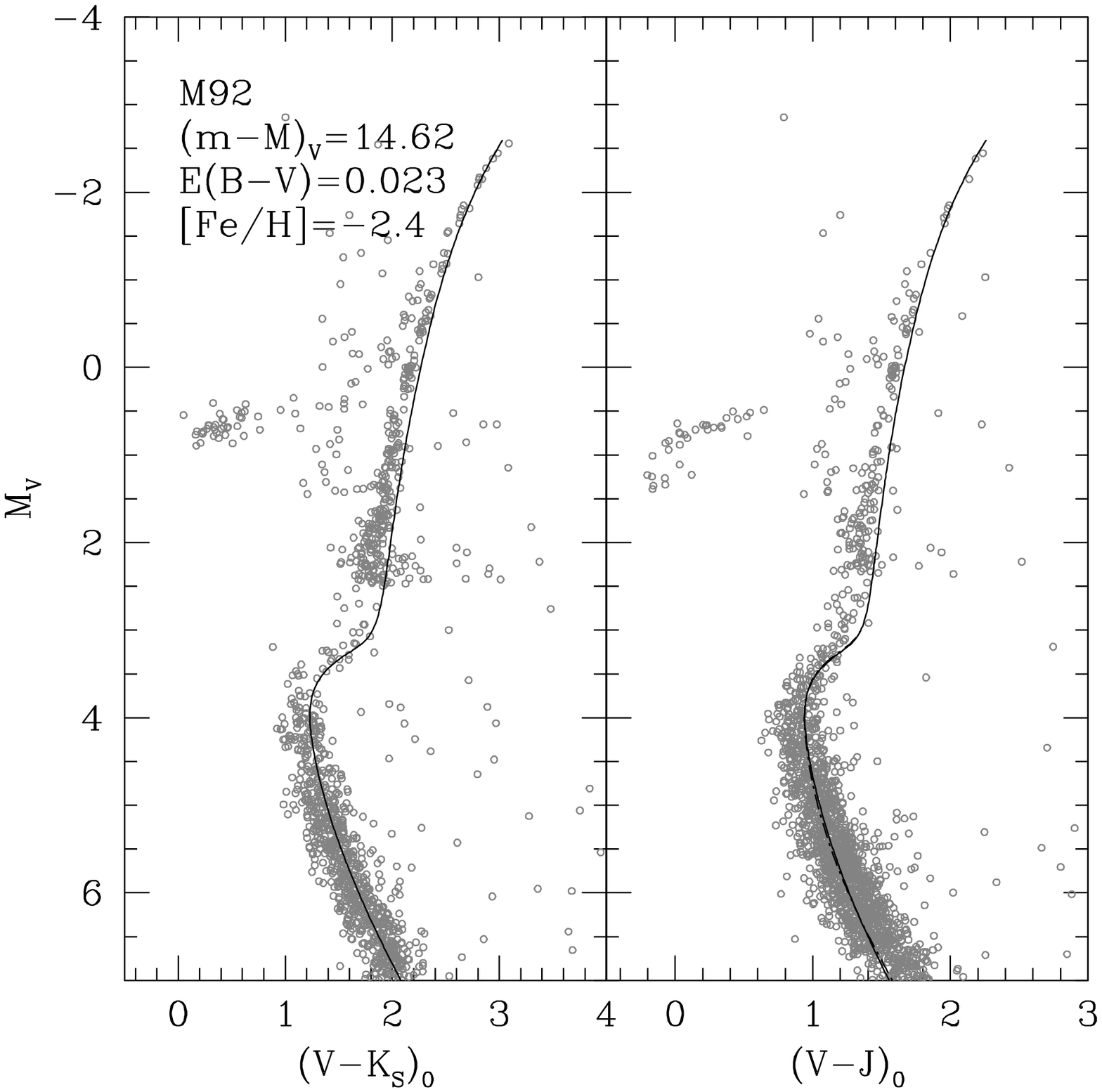}
\caption{As in the previous figure, except that the (same) isochrone is 
compared with our observations of M$\,$92.}
\label{M92_marcs}
\end{figure}

As shown by VCS10, a MARCS-transformed isochrone for 13.5 Gyr and
[Fe/H] $= -2.4$ provides a reasonably good fit to both the $BV$ and
$VI$ photometry of M$\,$92 --- aside from the fact that a small
colour shift must be applied to the $V-I$ colours in order to obtain
consistent fits to the various CMDs that can be generated from
$BV(RI)_C$ photometry.  Their additional observation that the isochrones
tend to deviate to the blue side of the observed lower MS is apparently
seen in especially the $[(V-J)_0,\,M_V]$-diagrams of M$\,$15 and
M$\,$92 as well (see Figures~\ref{M15_marcs} and~\ref{M92_marcs}).  We
have no explanation for such deviations, which appear to be most
pronounced in the $VJ$ photometry for M$\,$15.  

Perhaps the most serious concern is that, in both the $V-J$ and $V-K_S$
colour planes, the isochrones lie consistently to the red of the observed
RGBs of M$\,$15 and M$\,$92.  Near the base of the giant branch, the
offsets amount to $\sim 0.12$ mag in $V-K_S$ and $\sim 0.09$ mag in $V-J$ and the typical
error in the photometry ($\sim 0.02$ mag) cannot explain these colour offsets.  Surprisingly, VCS10 find that the same isochrones
provide a good fit to optical CMDs, apart from the small, constant offset
between the predicted and observed RGB on the $[(V-I)_0,\,M_V]$-plane previously mentioned.
Since we are using the same $V$-band photometry as VCS10, one might
question the accuracy of the $J$ and $K_S$ photometry for M$\,$15 and
M$\,$92.  However, this is not a viable explanation because the near-IR
photometry for the cluster giants, as with all our CFHT clusters, comes
directly from 2MASS, and such a large zero-point offset is highly unlikely.  Indeed, because the isochrones
reproduce $BV(RI)_C$ observations quite well, one wonders whether there
is a problem with the synthetic $J$ and $K_S$ magnitudes that are
derived from MARCS model atmospheres, though we have no other reason to
question to these results.  Further work is clearly needed to understand
this puzzle.

\subsection{Field Subdwarfs ($-$2.2 $\lesssim$ [Fe/H] $\lesssim -$ 0.5)}
\label{sec:subdwarftests}

\begin{figure*}
\includegraphics[bb= 0 0 446 700, width=10cm, angle=270]{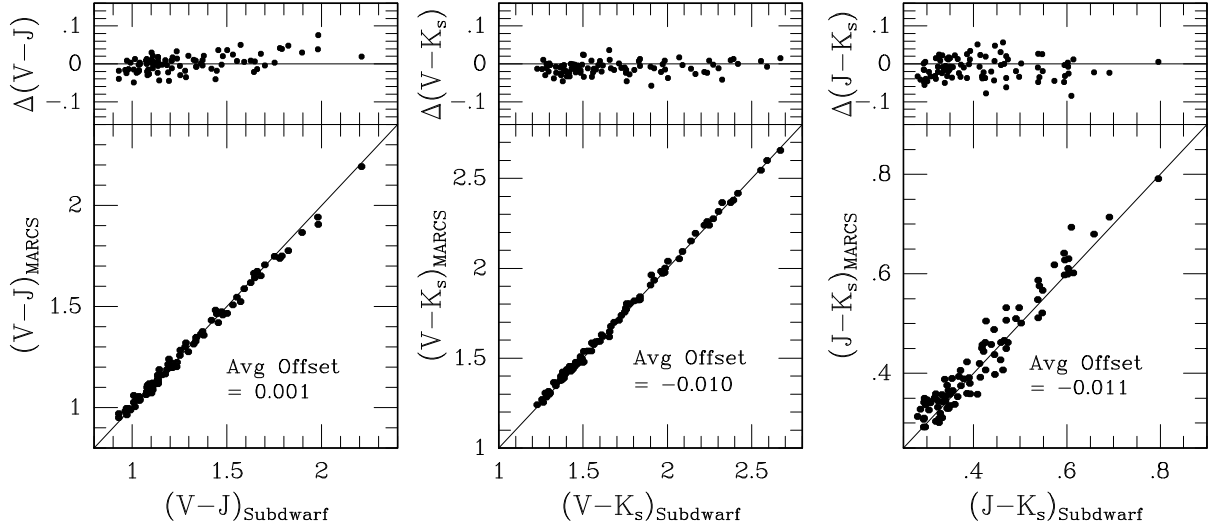}
\caption{ Comparison of the observed subdwarf colours with those
predicted by MARCS colour transformations on the assumption of the
$\Teff$, $\log\,g$ and [Fe/H] determinations of CRMBA. The average
offsets noted in each panel are relative to the observed colours. The upper panels plot the differences in the colours as a function of the subdwarf colors in the sense {\it observed minus
predicted}.}
\label{CG_V_K_2mass}
\end{figure*}

Subdwarfs are metal-poor stars with halo kinematics whose orbits have
brought them into the solar neighbourhood.  With many of these
stars having accurate distances determinations from {\it Hipparcos}
and recent estimates of $\log\,g$, and [Fe/H] based on
high-resolution spectroscopy, they provide strong constraints on the
$\Teff$ scale of isochrones and colour-$\Teff$ relations at low
metallicities. In the following analysis, we use a sample of $\sim
100$ subdwarfs compiled by VCS10. The photometric and fundamental
stellar properties (e.g., $\Teff$, $log\,g$ and [Fe/H]) for the majority
of these stars were taken from Table 8 of CRMBA with the addition of 4
stars from \citet{Clem1999}. For a further description on the
properties and choice of this sample, see VCS10.

In figure \ref{CG_V_K_2mass} we show the level of agreement between
the observed and predicted 2MASS colours, ($V-K_S$), ($V-J$), and
($J-K_S$), of the subdwarfs in our sample.  The predicted MARCS
colours for each star are found by interpolating in tables of synthetic
colours assuming the $\Teff$, $\log\,g$, and [Fe/H] values given by CRMBA.
For these colour-colour comparisons we use only MARCS transformations
since the CRMBA relations are themselves based on most of the same
stars used here, so they will necessarily yield colours that agree
well, in the mean, with the observed photometry.  As noted in the three
panels of figure \ref{CG_V_K_2mass}, offsets of $\sim$0.01 mag to the
blue for the predicted ($V-K_S$) and ($J-K_S$) are needed to achieve
consistency with observations.  The predicted ($V-J$) colours show a
very small mean offset of 0.001 mag from the observed colours, though
the largest deviations occur for the reddest stars, which suggests
(perhaps) that the transformations to $V-J$ for cooler dwarf stars
may need some adjustment.  (Recall that, although this evidence is
based on very few cool subdwarfs with measured $V-J$ colours, similar
problems were seen in our comparisons with the $V-J$ photometry for
MS stars in the low metallicity clusters M$\,$92, M$\,$15, M$\,$13
and M$\,$5; see the previous subsections.)

While the comparisons described above are completely independent of
stellar evolutionary models, instructive comparisons of the theoretical
isochrones with the {\it Hipparcos} subdwarfs can also be carried out.
By superimposing subdwarfs onto a grid of isochrones for a wide range in
[Fe/H] (and a fixed age --- though isochrones exhibit essentially no age
dependence at faint luminosities), it is possible to examine the
consistency of the observed [Fe/H] values and colours with those inferred
from the isochrones, given the observed temperatures.   If a star lies on
the same isochrone (say, one with [Fe/H]= $-$1.5) in all of the colour
planes, one can conclude that the colour-$\Teff$ transformations are
consistent for those colours.  If any obvious discrepancies exist for
different filter combinations, then this is an indication of a problem
with the colour-temperature relations for one or more colour indices (i.e.,
they do not adequately represent the spectrum of a star in the
wavelength ranges spanned by the relevant filters).

\begin{figure*}
\includegraphics[width=10cm, angle=270]{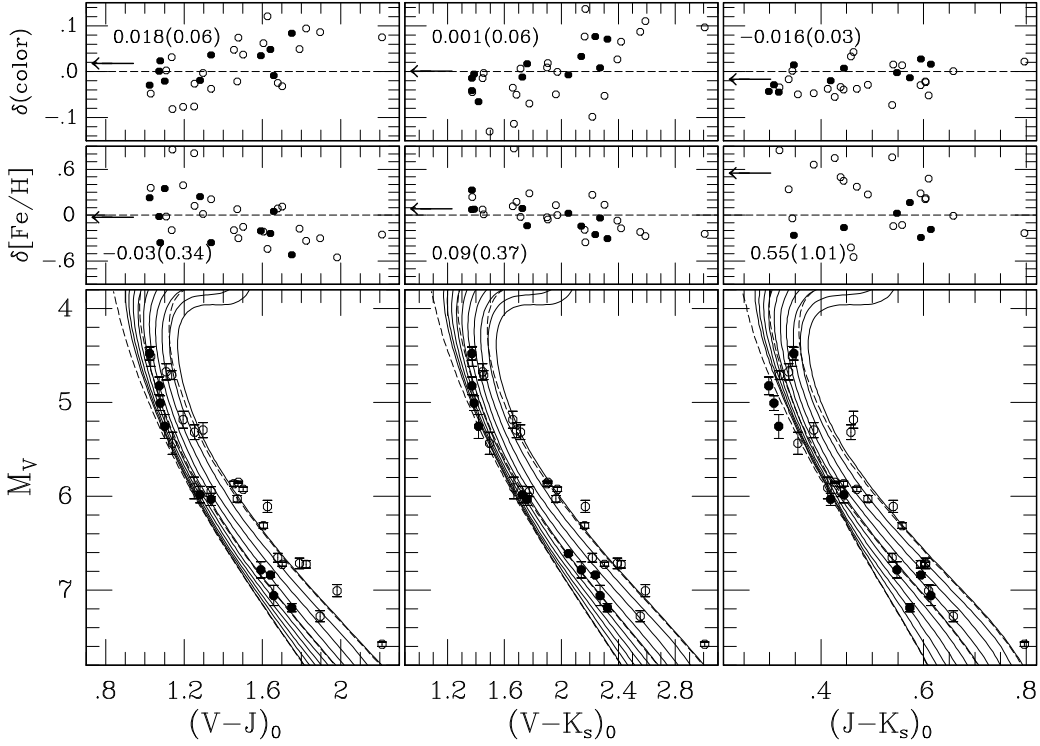}
\caption{{\it Lower Panels}:  Sample of Hipparcos subdwarfs overlying 12
Gyr Victoria isochrones in the $[(V-J)_0,\,M_V]$-, $[(V-K_S)_0,\,M_V]$- and
$[(J-K_S)_0,\,M_V]$-planes. The isochrones plotted here have [Fe/H] values
of $-2.4$ to $-0.6$ in increments of 0.2 dex, from left to right, and
MARCS transformations have been used to transpose the models to the various
CMDs. To demonstrate the age insensitivity for magnitudes $M_V\geq 4.5$, dashed lines show 10 Gyr isochrones for [Fe/H] = -0.6, -1.4, and -2.4.
The photometry and metallicity of each subdwarf is taken from
CRMBA, and the error bars represent 1$\sigma$ uncertainties in the
$M_V$ values as derived from {\it Hipparcos} parallaxes.   Subdwarfs
having [Fe/H]$\leq -1.2$, as given by CRMBA, are plotted as filled
circles, and those with higher metallicities are plotted as open circles.
{\it Middle Panels}:   Plotted as a function of colour, the difference
between the CRMBA estimate of [Fe/H] for each star and that inferred
from the interpolated (or extrapolated when necessary) isochrone that
matches its location on the colour--$M_V$ diagram in the lower panel. The arrows and the
numbers, together with the standard deviation (in parentheses), indicate
the mean $\delta$[Fe/H] for all of the subdwarfs that have been considered.. 
{\it Upper Panels}:  The difference in colour that would need to be
applied to each subdwarf in order to achieve perfect consistency of
its position relative to the isochrones in the lower panel. n this case, the arrows and numbers indicate the
mean values of $\delta$(colour) for the subdwarf sample.}
\label{subdwarfs_marcs}
\end{figure*}

\begin{figure*}
\includegraphics[width=10cm,angle=270]{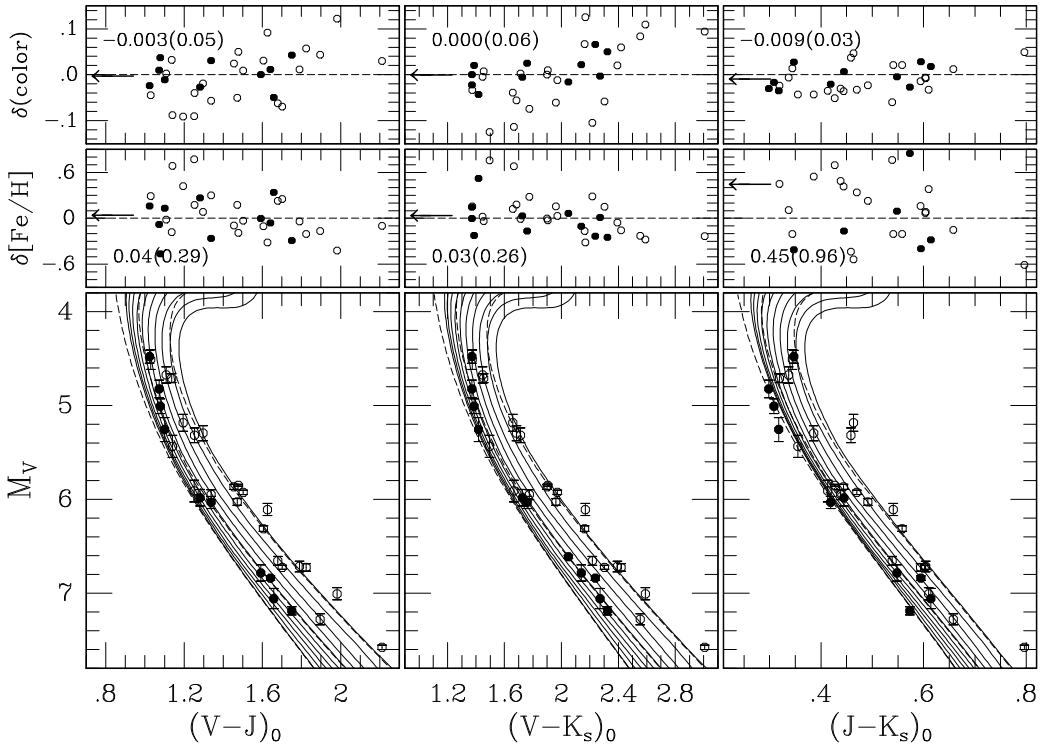}
\caption{Similar to previous figure, except that the isochrones are
transformed using the colour--$\Teff$ relations developed by CRMBA.}
\label{subdwarfs_luca}
\end{figure*}

In the lower panel of figure \ref{subdwarfs_marcs}, we superimpose
{\it Hipparcos} (\citet{Hipparcos}) subdwarfs onto 12 Gyr Victoria isochrones for [Fe/H]
values from $-2.4$ to $-0.6$ (in the direction from left to right)
transformed into the $[(V-J)\,M_V]$-, $[(V-K_S,\,M_V]$- and
$[(J-K_S),\,M_V]$-planes using the MARCS transformations.  For each
subdwarf, the CRMBA estimate of its [Fe/H] value minus the [Fe/H] value 
of the isochrone on which the subdwarf is located is plotted in the
middle panel, while the upper panel plots the difference between the
observed subdwarf colour and that of the isochrone which has the same
metallicity as the star (at the same $M_V$).  These panels provide a
measure of the quality of the fit to the observations.  For example, if
the isochrones were too red, most of the points would have positive
$\delta$[Fe/H] values and negative $\delta$(colour) values.  
Figures similar to  \ref{subdwarfs_marcs} and \ref{subdwarfs_luca}, but for the $(B-V)$, $(V-R)$, and $(V-I)$
colour planes, are given in the study by VCS10, where the individual subdwarfs are
identified.


The subdwarfs in the $V-K_S$ and $V-J$ colour planes show significant
relative shifts with respect to the isochrones (i.e., the subdwarfs
do not lie on the same position relative to the same isochrone in both
colours).  However, the scatter on the $\delta$[Fe/H] and $\delta$(colour)
planes is quite symmetric overall in both $V-K_S$ and $V-J$, with most of
the redder stars having negative $\delta$[Fe/H] and positive $\delta$(colour)
values, whereas most of the bluer stars have positive $\delta$[Fe/H]
and negative $\delta$(colour) values.  Thus, the isochrones are consistent
with the  observations, in the mean, with only a slight trend
in the sense that the isochrones are somewhat too blue for the redder
stars and too red for the bluer stars.  

A different conclusion is drawn regarding the $J-K_S$ transformations shown
in the right panel of Figure \ref{subdwarfs_marcs}; the MARCS-transformed
isochrones imply implausibly low values of [Fe/H], which are contrary to the
implications from the other colour planes.  When looking at the $\delta$[Fe/H]
and $\delta$(colour) plots in the upper panels, one finds that most of the
points have positive $\delta$[Fe/H] and negative $\delta$(colour) values,
indicating that the isochrones are too red, in the mean, relative to the
observations. The $J-K_S$ panels also indicate that there is no obvious
trend with colour or with metallicity.

Figure \ref{subdwarfs_luca} plots the same isochrones as in
Figure \ref{subdwarfs_marcs}, but transformed to the various colour planes
using the CRMBA colour--$\Teff$ relations.  When comparing the subdwarf
colours and metallicities with those of the isochrones, we find there is
no obvious trend with either colour or metallicity, which indicates that the
CRMBA-transformed isochrones agree well, in the mean, with the observations,
while the absence of trends provides a verification that the functional form used 
in CRMBA transformations well 
represents the (bulk of) stars upon which it is built.

\section{Summary}
\label{sec:summary}

\begin{figure}
\includegraphics[bb= 0 140 446 700, width=7cm]{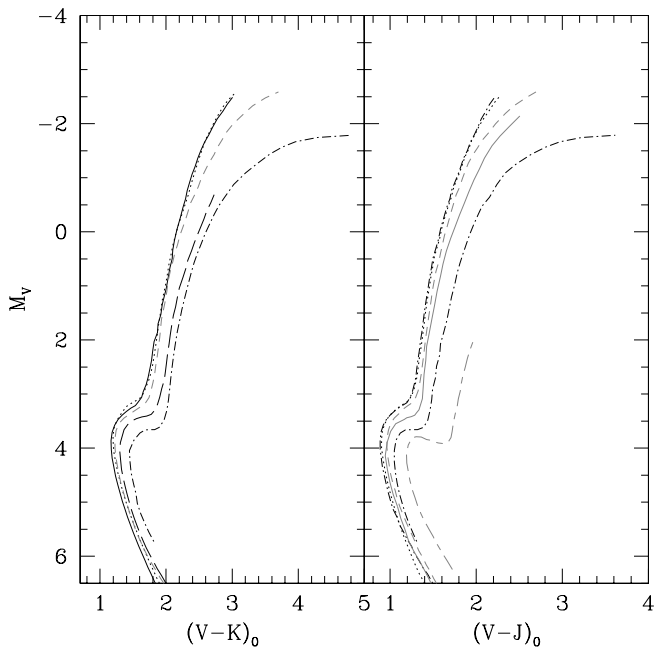}
\caption{ Our cluster fiducials mapped to the $[(V-K_S)_0,\,M_V]$- and
$[(V-J)_0,\,M_V]$-planes using the reddenings and distance moduli given in
Table \ref{cluster_params}.  From left to right, corresponding to
increasing metallicity, are M$\,$15 (solid black), M$\,$92 (dotted
black), M$\,$13 (grey short-dashed), NGC$\,$1851 (black long-dashed), M$\,$5 (black solid line), M$\,$71 (black dot-dashed) and NGC$\,$6791 (grey long-short dashed).
}
\label{fiducials}
\end{figure}

We have derived fiducial sequences for the Galactic star clusters
M$\,$15, M$\,$92, M$\,$13, M$\,$5, NGC$\,$1851, M$\,$71, and NGC 6791
(encompassing $-2.4 \lesssim \textnormal{[Fe/H]} \lesssim +0.3$) using
near-IR observations obtained with the WIRCam imager on the
Canada-France-Hawaii Telescope and the HAWK-I detector on the VLT.
These fiducial sequences, which are presented in Figure~\ref{fiducials}
on the $[(V-J)_0,\,M_V]$- and $[(V-K_S)_0,\,M_V]$-planes, illustrate how
the photometric properties of these old stellar populations vary as a
function of [Fe/H].  With spectroscopic metallicity determinations and
relative age estimates that are accurate to within $\pm 0.25$ dex and $\pm 1.5$--2
Gyr, respectively, these fiducials can, in principle, be used to
photometrically determine the age and metallicity of resolved stellar
systems. Unlike isochrone analyses, metallicity determinations made
through comparisons with fiducials are independent of stellar
evolutionary models. Hence, these fiducials provide a set of empirical
isochrones that can serve as valuable tools for future stellar
population investigations involving the 2MASS filters.

\vspace{0.27cm}

Based on these data, a summary of our results is as follows:

\vspace{0.27cm}

(1) The predicted $V-J$, $V-K_S$, and $J-K_S$ colours given by the MARCS
transformations for local subdwarfs agree well with those observed
(i.e., to within $\sim 0.01$ mag) when the temperatures and metallicities
of the latter are as given by CRMBA.

\vspace{0.27cm}

(2) While isochrones appear to be too blue when compared with lower
main sequence stars in metal-poor GCs on the $[(V-J)_0,\,M_V]$-plane,
no significant problems are found matching their $V-K_S$ colours.

\vspace{0.27cm}

(3) Our analysis of cluster RGB stars indicates that the isochrones 
and the colour transformations faithfully reproduce the properties of
metal-rich giants, but not those of lower metallicity. In fact, the
RGB segments of the isochrones become systematically redder than the
observed RGBs with decreasing [Fe/H].  It seems unlikely that errors
in the model $\Teff$ scale are sufficient to cause this problem, 
because the discrepancies are primarily on the $V-J$ and $V-K_S$
colour planes.  The same MARCS-transformed isochrones are able to
reproduce the CMDs of M$\,$92 derived from $BV(RI)_C$ photometry quite
well.  This seems to suggest that there are problems with the predicted
$J$ and $K_S$ magnitudes at low metallicities, but we are not able to
provide an explanation for the origin of such difficulties.

\acknowledgements
This work has been supported by the Natural Sciences and Engineering
Research Council of Canada through a Discovery Grant to DAV. We are
grateful to the anonymous referee for valuable comments. Additionally, CMB 
thanks Loic Albert and Aaron Dotter for insightful
discussions which improved both the observations and analysis in this
paper.


\begin{table*}
\label{NGC6791_table}
 \hspace{0.033\textwidth}
 \begin{minipage}{0.4\textwidth}
 \centering
\begin{tabular}{l  l} \hline \hline
$V$ & $V-J$    \\
  \hline
19.817 & 2.062 \\
19.659 & 2.001 \\
19.515 & 1.944 \\
19.363 & 1.889 \\
19.215 & 1.840 \\
19.067 & 1.787\\
18.924 & 1.738\\
18.771 & 1.692\\
18.633 & 1.656\\
18.486 & 1.625\\
18.331 & 1.595 \\
18.172 & 1.567 \\
18.011 & 1.548 \\
17.845 & 1.529 \\
17.676 & 1.528 \\
17.516 & 1.548 \\
17.388 & 1.588 \\
17.357 & 1.639 \\
17.367 & 1.699 \\
\hline
 \end{tabular}

 \end{minipage}
 \hspace{0.033\textwidth}
 \begin{minipage}{0.4\textwidth}
 \centering
\begin{tabular}{ l l} \hline \hline
$V$ & $V-J$    \\
\hline17.411 & 1.762 \\
17.437 & 1.831 \\
17.476 & 1.897 \\
17.485 & 1.953 \\
17.459 & 2.011\\
17.376 & 2.043 \\
17.263 & 2.058 \\
17.109 & 2.071\\
16.959 & 2.091\\
16.809 & 2.114\\
16.659 & 2.130\\
16.509 & 2.152\\
16.359 & 2.164\\
16.209 & 2.186\\
16.059 & 2.211\\
15.909 & 2.236\\
15.759 & 2.264\\
15.609 & 2.299\\
 & \\
\hline
 \end{tabular}
 \end{minipage}
\caption[Fiducial sequence for NGC$\,$6791.] {Fiducial sequence for
the open cluster NGC$\,$6791 as shown in Figure
\ref{NGC6791_fiducial}.}
\end{table*}

\begin{table*}
 \hspace{0.033\textwidth}
 \begin{minipage}{0.4\textwidth}
 \centering
\begin{tabular}{l l l} \hline \hline
$V$ & $V-J$ & $V-K_S$   \\
  \hline
19.505&1.805&2.408\\
19.338&1.757&2.338\\
19.168&1.712&2.273\\
18.991&1.670&2.214\\
18.807&1.635&2.163\\
18.622&1.604&2.138\\
18.429&1.576&2.099\\
18.232&1.555&2.068\\
18.029&1.542&2.051\\
17.824&1.542&2.040\\
17.634&1.563&2.110\\
17.500&1.611&2.199\\
17.445&1.673&2.278\\
17.439&1.745&2.353\\
17.441&1.814&2.424\\
17.396&1.882&2.492\\
17.267&1.931&2.564\\
17.084&1.959&2.606\\
16.885&1.977&2.631\\
16.688&1.989&2.649\\
16.485&2.021&2.667\\
16.268&2.035&2.687\\
16.060&2.070&2.709\\
\hline
 \end{tabular}

 \end{minipage}
 \hspace{0.033\textwidth}
 \begin{minipage}{0.4\textwidth}
 \centering
\begin{tabular}{l l l} \hline \hline
$V$ & $V-J$ & $V-K_S$   \\
\hline
15.854&2.080&2.734\\
15.627&2.120&2.766\\
15.426&2.152&2.798\\
15.217&2.177&2.835\\
14.986&2.209&2.880\\
14.736&2.247&2.935\\
14.538&2.274&2.974\\
14.356&2.306&3.020\\
14.177&2.342&3.071\\
13.993&2.381&3.129\\
13.818&2.422&3.187\\
13.646&2.465&3.250\\
13.459&2.516&3.324\\
13.305&2.562&3.390\\
13.139&2.648&3.481\\
12.906&2.733*&3.619*\\
12.693&2.833*&3.800*\\
12.523&2.962*&3.970*\\
12.347&3.137*&4.164*\\
12.212&3.300*&4.350*\\
12.116&3.469*&4.552*\\
12.035&3.717*&4.883*\\
11.997&4.109*&5.401*\\
\hline
 \end{tabular}
 \end{minipage}
\caption[Fiducial sequences for M$\,$71.] {Fiducial sequences for the
globular cluster M$\,$71 as shown in Figure \ref{M71_fiducial}. Due to
the sparseness of the RGB, those points with $V <$13 are marked by asterisks to indicate
they have large uncertainties associated with them.}
\end{table*}

\begin{table*}
 \hspace{0.033\textwidth}
 \begin{minipage}{0.4\textwidth}
 \centering
\begin{tabular}{l l } \hline \hline
$V$ & ($V-J$)   \\
  \hline
21.95&  2.01\\
21.84&  1.96\\
21.60&  1.84\\
21.37&  1.74\\
21.16&  1.65\\
20.95&  1.56\\
20.75&  1.50\\
20.55&  1.43\\
20.36&  1.36\\
20.16&  1.30\\
19.96&  1.26\\
19.75&  1.21\\
19.53&  1.17\\
19.31&  1.12\\
19.07&  1.08\\
18.83&  1.05\\
18.59&  1.03\\
18.34&  1.05\\
18.14&  1.09\\
18.00&  1.15\\
17.92&  1.25\\
17.89&  1.28\\
17.87&  1.33\\
17.84&  1.37\\
17.80&  1.39\\
\hline
 \end{tabular}
 \end{minipage}
 \hspace{0.033\textwidth}
 \begin{minipage}{0.4\textwidth}
 \centering
\begin{tabular}{l l} \hline \hline
$V$ & ($V-J$)   \\
\hline
17.74&  1.42\\
17.54&  1.46\\
17.32&  1.47\\
17.08&  1.48\\
16.83&  1.50\\
16.57&  1.51\\
16.33&  1.54\\
16.05&  1.57\\
15.81&  1.60\\
15.55&  1.63\\
15.27&  1.67\\
15.04&  1.70\\
14.80&  1.74\\
14.57&  1.79\\
14.35&  1.84\\
14.14&  1.89\\
13.91&  1.95\\
13.71&  2.00\\
13.49&  2.06\\
13.29&  2.12\\
13.09&  2.18\\
12.86&  2.27\\
12.68&  2.36\\
12.50&  2.47\\
12.31&  2.59\\
\hline
 \end{tabular}
 \end{minipage}
\caption[Fiducial sequence for M$\,$5.] {Fiducial sequence for the
globular cluster M$\,$5 as shown in Figure \ref{M5_fiducial}.}
\end{table*}

\begin{table*}
 \hspace{0.033\textwidth}
 \begin{minipage}{0.4\textwidth}
 \centering
\begin{tabular}{l l } \hline \hline
$V$ & $V-K_S$   \\
  \hline
21.993 &2.087\\
21.856 &2.020\\
21.729 &1.959\\
21.609 &1.904\\
21.473 &1.844\\
21.344 &1.789\\
21.212 &1.741\\
21.077 &1.696\\
20.938 &1.652\\
20.797 &1.611\\
20.655 &1.573\\
20.510 &1.536\\
20.360 &1.502\\
20.212 &1.472\\
20.052 &1.444\\
19.893 &1.419\\
19.731 &1.402\\
19.564 &1.393\\
19.399 &1.400\\
19.159 &1.457\\
19.063 &1.511\\
18.996 &1.575\\
18.952 &1.644\\
18.923 &1.725\\
18.906 &1.797\\
18.876 &1.868\\
\hline
 \end{tabular}
 \end{minipage}
 \hspace{0.033\textwidth}
 \begin{minipage}{0.4\textwidth}
 \centering
\begin{tabular}{l l} \hline \hline
$V$ & $V-K_S$   \\
\hline
18.806 &1.914\\
18.695 &1.963\\
18.557 &1.998\\
18.401 &2.024\\
18.245 &2.064\\
18.083 &2.081\\
17.917 &2.099\\
17.740 &2.117\\
17.570&2.136\\
17.410 &2.155\\
17.237 &2.177\\
17.050 &2.202\\
16.885 &2.237\\
16.720 &2.262\\
16.540 &2.292\\
16.342 &2.348\\
16.180 &2.389\\
16.027 &2.420\\
15.857 &2.475\\
15.702 &2.519\\
15.558 &2.568\\
15.403 &2.617\\
15.262 &2.665\\
15.113 &2.708\\
14.964 &2.768\\
14.815 &2.818\\
\hline
 \end{tabular}
 \end{minipage}
\caption[Fiducial sequence for NGC$\,$1851.] {Fiducial sequence for
the globular cluster NGC$\,$1851 as shown in Figure
\ref{NGC1851_fiducial}.}
\end{table*}

\begin{table*}
 \hspace{0.033\textwidth}
 \begin{minipage}{0.4\textwidth}
 \centering
\begin{tabular}{l l l} \hline \hline
$V$ & $V-J$ & $V-K_S$   \\
  \hline
22.039&2.011&2.606\\
21.884&1.939&2.513\\
21.740&1.883&2.416\\
21.606&1.832&2.344\\
21.460&1.766&2.267\\
21.306&1.709&2.176\\
21.173&1.653&2.120\\
21.030&1.596&2.042\\
20.891&1.541&1.988\\
20.761&1.501&1.931\\
20.640&1.455&1.868\\
20.502&1.414&1.801\\
20.372&1.368&1.760\\
20.249&1.334&1.712\\
20.112&1.299&1.662\\
19.979&1.260&1.620\\
19.839&1.220&1.578\\
19.700&1.182&1.518\\
19.560&1.146&1.491\\
19.411&1.119&1.454\\
19.263&1.095&1.419\\
19.113&1.063&1.397\\
18.962&1.042&1.369\\
18.800&1.024&1.342\\
18.637&1.008&1.320\\
18.472&0.999&1.327\\
18.306&0.999&1.338\\
18.152&1.014&1.368\\
18.022&1.043&1.430\\
17.923&1.083*&1.506*\\
17.842&1.139*&1.556*\\
17.751&1.192*&1.632*\\
17.692&1.242*&1.684*\\
17.642&1.290*&1.734*\\
\hline
 \end{tabular}

 \end{minipage}
 \hspace{0.033\textwidth}
 \begin{minipage}{0.4\textwidth}
 \centering
\begin{tabular}{l l l} \hline \hline
$V$ & $V-J$ & $V-K_S$   \\
\hline
17.548&1.331*&1.782*\\
17.420&1.360*&1.814*\\
17.271&1.381*&1.844*\\
17.112&1.397&1.868\\
16.952&1.411&1.888\\
16.786&1.424&1.907\\
16.433&1.451&1.945\\
16.104&1.478&1.974\\
15.747&1.511&2.031\\
15.585&1.527&2.050\\
15.408&1.546&2.072\\
15.214&1.569&2.104\\
15.036&1.601&2.145\\
14.880&1.620&2.173\\
14.715&1.643&2.215\\
14.546&1.667&2.249\\
14.393&1.697&2.298\\
14.238&1.722&2.342\\
14.088&1.748&2.379\\
13.945&1.774&2.415\\
13.800&1.802&2.475\\
13.646&1.843&2.529\\
13.499&1.875&2.573\\
13.360&1.907&2.628\\
13.211&1.944&2.689\\
13.051&1.999&2.768\\
12.923&2.053&2.818\\
12.787&2.104&2.880\\
12.642&2.182&2.952\\
12.487&2.258&3.058\\
12.289&2.373&3.223\\
12.152&2.474&3.352\\
12.005&2.612&3.547\\
11.895&2.729&3.741\\
\hline
 \end{tabular}
 \end{minipage}
\caption[Fiducial sequences for M$\,$13.] {Fiducial sequences for the
globular cluster M$\,$13 as shown in Figure \ref{M13_fiducial}. Due to
the lack of sufficient data at some magnitudes, isochrones were used
to determine the fiducial points indicated by asterisks. Therefore
these points have large uncertainties associated with them.}
\end{table*}

\begin{table*}
 \hspace{0.033\textwidth}
 \begin{minipage}{0.4\textwidth}
 \centering
\begin{tabular}{l l l} \hline \hline
$V$ & $V-J$ & $V-K_S$   \\
  \hline
21.994&1.833&2.309\\
21.836&1.752&2.241\\
21.688&1.684&2.177\\
21.550&1.621&2.119\\
21.420&1.570&2.064\\
21.187&1.474&1.969\\
21.064&1.433&1.920\\
20.924&1.384&1.867\\
20.792&1.347&1.818\\
20.667&1.314&1.772\\
20.537&1.280&1.725\\
20.403&1.247&1.678\\
20.273&1.216&1.635\\
20.142&1.186&1.593\\
20.000&1.155&1.549\\
19.859&1.127&1.509\\
19.716&1.099&1.470\\
19.572&1.073&1.432\\
19.422&1.047&1.396\\
19.267&1.023&1.361\\
19.105&1.000&1.328\\
18.938&0.980&1.300\\
18.780&0.965&1.278\\
18.617&0.953&1.261\\
18.451&0.951&1.259\\
18.294&0.963&1.276\\
18.164&0.990*&1.314*\\
18.062&1.026*&1.365*\\
17.981&1.068*&1.415*\\
17.912&1.113*&1.450*\\
17.852&1.163*&1.521*\\
17.803&1.210*&1.589*\\
17.751&1.261*&1.662*\\
17.652&1.300*&1.723*\\
\hline
 \end{tabular}

 \end{minipage}
 \hspace{0.033\textwidth}
 \begin{minipage}{0.4\textwidth}
 \centering
\begin{tabular}{l l l} \hline \hline
$V$ & $V-J$ & $V-K_S$   \\
\hline

17.541&1.326*&1.758*\\
17.418&1.346*&1.800*\\
17.279&1.368*&1.832*\\
17.130&1.376*&1.857*\\
16.970&1.391*&1.880*\\
16.801&1.406*&1.901*\\
16.623&1.421*&1.922*\\
16.453&1.435*&1.942*\\
16.292&1.448&1.961\\
16.118&1.463&1.982\\
15.929&1.479&2.005\\
15.637&1.507&2.054\\
15.456&1.525&2.080\\
15.256&1.547&2.109\\
15.089&1.565&2.136\\
14.925&1.595&2.162\\
14.744&1.627&2.201\\
14.541&1.653&2.247\\
14.321&1.692&2.308\\
14.140&1.727&2.357\\
13.984&1.761&2.399\\
13.827&1.783&2.433\\
13.682&1.818&2.467\\
13.528&1.857&2.505\\
13.379&1.889&2.544\\
13.236&1.920&2.583\\
13.085&1.964&2.626\\
12.933&1.999&2.675\\
12.798&2.046&2.730\\
12.655&2.086&2.787\\
12.503&2.147&2.854\\
12.362&2.208&2.922\\
12.235&2.262&2.998\\
12.103&2.332&3.108\\
\hline
 \end{tabular}
 \end{minipage}
\caption[Fiducial sequences for M$\,$92.] {Fiducial sequences for the
globular cluster M$\,$92 as shown in Figure \ref{M92_fiducial}. Due to
the lack of sufficient data at some magnitudes, isochrones were used
to determine the fiducial points indicated by asterisks. Therefore
these points have large uncertainties associated with them.}
\end{table*}

\begin{table*}
 \hspace{0.033\textwidth}
 \begin{minipage}{0.4\textwidth}
 \centering
\begin{tabular}{l l l} \hline \hline
$V$ & $V-J$ & $V-K_S$   \\
  \hline
22.010&1.764&2.140\\
21.887&1.694&2.079\\
21.764&1.653&2.030\\
21.624&1.604&1.977\\
21.492&1.547&1.928\\
21.367&1.504&1.882\\
21.237&1.450&1.835\\
21.103&1.417&1.788\\
20.973&1.366&1.745\\
20.842&1.326&1.703\\
20.700&1.277&1.659\\
20.559&1.247&1.619\\
20.416&1.199&1.580\\
20.272&1.173&1.542\\
20.122&1.147&1.506\\
19.967&1.123&1.471\\
19.805&1.100&1.438\\
19.638&1.080&1.410\\
19.480&1.065&1.388\\
19.317&1.053&1.371\\
19.151&1.051&1.369\\
18.994&1.063&1.386\\
18.864&1.090&1.424\\
18.762&1.126&1.475\\
18.681&1.168*&1.535\\
18.612&1.213*&1.600\\
18.552&1.263*&1.671\\
18.503&1.310*&1.739\\
18.451&1.361*&1.772\\
18.352&1.400*&1.833\\
18.241&1.426*&1.858*\\
18.118&1.446*&1.900*\\
\hline
 \end{tabular}

 \end{minipage}
 \hspace{0.033\textwidth}
 \begin{minipage}{0.4\textwidth}
 \centering
\begin{tabular}{l l l} \hline \hline
$V$ & $V-J$ & $V-K_S$   \\
\hline
17.979&1.468*&1.932*\\
17.830&1.476*&1.957*\\
17.670&1.491*&1.980*\\
17.501&1.506*&2.001*\\
17.323&1.521&2.022\\
17.153&1.535&2.072\\
16.992&1.548&2.081\\
16.818&1.563&2.122\\
16.629&1.579&2.145\\
16.337&1.607&2.204\\
16.156&1.625&2.230\\
15.956&1.647&2.259\\
15.789&1.665&2.296\\
15.625&1.695&2.302\\
15.444&1.727&2.331\\
15.241&1.753&2.367\\
15.021&1.792&2.418\\
14.840&1.827&2.457\\
14.684&1.861&2.499\\
14.527&1.883&2.523\\
14.382&1.918&2.557\\
14.228&1.957&2.595\\
14.079&1.989&2.644\\
13.936&2.020&2.683\\
13.785&2.064&2.726\\
13.633&2.099&2.785\\
13.498&2.146&2.850\\
13.355&2.186&2.907\\
13.203&2.227&2.976\\
13.062&2.278&3.062\\
12.935&2.322&3.128\\
12.803&2.382&3.208\\

\hline
 \end{tabular}
 \end{minipage}
\caption[Fiducial sequences for M$\,$15.] {Fiducial sequences for the
globular cluster M$\,$15 as shown in Figure \ref{M15_fiducial}. Due to
the lack of sufficient data at some magnitudes, isochrones were used
to determine the fiducial points indicated by asterisks. Therefore
these points have large uncertainties associated with them.}
\end{table*}

\end{document}